\documentclass[final,5p,times,twocolumn,authoryear]{elsarticle}

\usepackage[english]{babel}

\usepackage{amsmath,amssymb}
\usepackage{graphicx}
\usepackage{threeparttable}
\usepackage{multirow}
\usepackage{threeparttable}
\usepackage{bm}
\usepackage{tabularx}
\usepackage{soul}
\usepackage[table]{xcolor}
\usepackage{hyperref}
\usepackage{float}
\usepackage{caption}
\usepackage{array}

\newcolumntype{C}[1]{>{\centering\arraybackslash}m{#1}}

\journal{High Energy Astrophysics}

\begin{document}

\begin{frontmatter}



\title{Is Dark Energy Increasing or Decreasing in the Late Universe?}

\author[1]{Maryam Aghaei Abchouyeh}
\affiliation[1]{organization={Department of Physics and Astronomy, Sejong University},
            addressline={98 Gunja-Dong, Gwangjin-gu, Seoul},
            city={Seoul},
            postcode={143-747},
            country={South Korea}}

\author[1,2]{M. H. P. M. van Putten\corref{cor2}}
\cortext[cor2]{Corresponding author, mvp@sejong.ac.kr}

\affiliation[2]{organization={INAF-Osservatorio Astronomico di Capodimonte}, 
            addressline={Salita Moiariello 16},
            city={Napoli},
            postcode={I-80131},
            country={Italy}}

\begin{abstract}

The concordance cosmological model $\Lambda$CDM assumes dark energy to be a constant, consistent with early-time observations, evidenced by Planck-$\Lambda$CDM analyses. However, in the face of late-time tensions, the nature of dark energy remains a central open problem.
Modern precision cosmology offers a potential new window into the nature of dark energy in the $w(a){\rm CDM}$ framework, which provides a model-independent prescription for its unknown equation of state $w(a)$. 
A confrontation of $w(a){\rm CDM}$ with data generally constitutes a nonlinear inference problem. 
We find that $w_0w_a$ estimates  posterior to a fully non-linear $w(a)$ analysis are stabilized by the
{\it Baryon Acoustic Oscillation} (BAO) 
constraint on $c_M=\Omega_{m,0}h^2$, inherited for instance from Planck-$\Lambda$CDM analysis of the CMB. 
This implementation produces $w_0w_a$ estimates that are invariant under constraint-preserving variations in $\Omega_{m,0}$.
In contrast, the 
early-linearization $w_0w_a$CDM 
shows pronounced correlation with $\Omega_{m,0}$ even when preserving $c_M$. 
We quantify this correlation resulting from the non-commutativity of $w_0w_a$ estimation and linearization in $w(a)$CDM.
This discrepancy is demonstrated in controlled  mock-data experiments.
Applied to cosmic chronometer data, $w_0w_a$ estimates from correlation-free late linearization of $w(a)$CDM analysis favor $w_0<-1$, whereas $w_0w_a$CDM favors $w_0>-1$.
{If the correlation between $w_0$ and $\Omega_{m,0}$ in $w_0w_a$CDM is interpreted as arising from linearization effects rather than a physical origin, 
application to DESI DR2 may shift $w_0$ downward, potentially extending to $w_0 < -1$, corresponding to increasing dark energy at the present epoch. Alternatively, if the correlation is of hitherto unseen physical origin, the $w_0w_a$CDM parametrization is self-consistent and no such correction to the DESI inference {may be} required.
}
\end{abstract}



\end{frontmatter}


\section{Introduction}
\label{Sec:Intro}

Accelerated expansion of the late-time Universe \citep{Riess1998,Perl1998b,Perl1999} revived the nature of dark energy $\Lambda$ to the {frontier} of modern cosmology. The constant dark energy in $\Lambda$CDM successfully describes the \textit{Cosmic Microwave Background} (CMB) observed by \citeauthor{Aghanim2020} (\citeyear{Aghanim2020}). 

However, $\Lambda$CDM is notably challenged by independent precision measurements of the Hubble constant: $H_0 \simeq 73.04 \pm 1.04~\mathrm{km\,s^{-1}\,Mpc^{-1}}$ from the Local Distance Ladder (LDL), in significant tension with the Planck value $H_0 \simeq 67.36 \pm 0.54~\mathrm{km\,s^{-1}\,Mpc^{-1}}$ \citep{Aghanim2020,Ries2022}.

The $H_0$ tension has renewed interests to studies and observations of possible dynamical dark energy beyond $\Lambda$CDM, most recently {so} by DESI \citep{adame2025}. A common framework for studying this possibility is $w(a)$CDM, where $w(a)=p_{\mathrm{DE}}/\rho_{\mathrm{DE}}$denotes a general equation of state of dark energy as a function of the Friedmann scale factor $a=1/(1+z)$, or equivalently the cosmological redshift $z$ \citep{Huterer2001}. This is commonly implemented by the linearized formalism $w_0w_a$CDM defined by
\begin{equation}
    w(a)= w_0+(1-a)w_a.
\label{EQN_w0wa}
\end{equation}

\noindent The constants $w_0$ and $w_a$ appear in a projection of $w(a)$ onto a polynomial of degree one in this CPL 
formalism \citep{Chev2001,Lind2003}. 
A focal point in this probe for dynamical dark energy is 
$w_0$. Algebraically related to the deceleration parameter $q_0$, we have 

\begin{equation}
    w_0 < -1 \Longleftrightarrow q_0<q_0^{\Lambda},
    \label{EQN_w0}
\end{equation}

\noindent where $q_0^\Lambda\simeq -0.527$ refers to the Planck-$\Lambda$CDM estimate. 

\textcolor{black}{Based on $w_0w_a$CDM, DESI DR2 \citep{adame2024a} recently reported evidence for decreasing dark energy, with estimates $-1<w_0<0$ and $w_a<-1$ \citep{karim2025a}. This corresponds to thawing dark energy, with $(w_0,w_a)$ located in the fourth quadrant of the $w_0w_a$ plane centered on $(-1,0)$.
Unexpectedly, this result points to reverse inequalities in (\ref{EQN_w0}), in tension with other studies, particularly those involving the Local Distance Ladder (LDL) (e.g. \citealt{Cama2020,Simo2025}), even though DESI DR2 probes a very similar redshift range through six independent observatories covering $0.1 \lesssim z \lesssim 4$ \citep{karim2025c}. Since DESI measures BAO over $0.1<z\lesssim2.3$, including detailed corrections for spectral distortions from nonlinear evolution and low-redshift inhomogeneities in galaxy distributions, it is expected to probe dark energy in a manner broadly consistent with LDL-based analyses \citep{Lodha2025e,adame2025}.}

\textcolor{black}{The DESI analysis is based on (\ref{EQN_w0wa}), applying an upfront linearization of $w(a)$CDM. This approach inevitably carries systematic uncertainties through potential correlations in cosmological parameters such as the matter density $\Omega_{m,0}$, arising from the intrinsically nonlinear behavior of $w(a)$ beyond $\Lambda$CDM. Indeed, the applicability of (\ref{EQN_w0wa}) and related parametrizations has been questioned extensively for more than two decades (e.g. \cite{Huter2003,Ma2011,Feng2012,Pant2016,Colgin2021,Ness2025,Lee2025}), though without specifically identifying the underlying mathematical origin.}

Motivated by these discrepancies, 
we here focus on potential systematic
correlations resulting from linearization (\ref{EQN_w0wa}) upfront compared with linearization posterior to the fully nonlinear $w(a){\rm CDM}$ formalism with (\ref{EQN_w0wa}) --- {\it early} versus {\it late} linearization. {We highlight the crucial role of the BAO constraint in constrained and unconstrained measurement uncertainties, and their propagation in $w_0w_a$-estimates.}

Distinct behavior between early and late linearization is demonstrated through applications to mock data and to the cosmic chronometer data \citep{Faro2017}. While mock data provide a clean and controlled setting for studying the two approaches side-by-side and assessing potential systematics, cosmic chronometer data — a compilation of Hubble parameter measurements in the Local Universe ($0.07<z<1.95$) consistent with LDL and SH0ES \citep{Ries2022,Simo2025} — provide a suitable setting for illustrative estimates of $w_0w_a$, $H_0$, and $q_0$. Nevertheless, the distinct behavior observed in these analyses is representative and largely independent of the particular data set employed.

To begin, we discuss $w(a)$CDM and the stabilization of $w_0$ {estimates by the BAO constraint}, present in late but absent in early linearization (\S\ref{Sec:model}). Representative results are presented in \S\ref{Sec:Data}, including data preparation, linearization, and cosmological parameter estimation by Monte Carlo (MC) methods applied to mock and cosmic chronometer data. We interpret these results in light of (\ref{EQN_w0}) and comment on their significance for DESI in \S\ref{Sec:results}. 
We conclude our finding with implications for dynamical dark energy in \S\ref{Sec:conclusion}.

\section{BAO constrained equation of state of dark energy}
\label{Sec:model}

{$\Lambda$CDM with constant dark energy $\Lambda$ with equation of state parameter $w\equiv p_{\Lambda}/\rho_{\Lambda}=-1$ is} the most common cosmological model in describing the evolution of the cosmos. While it shows an excellent fit to Planck CMB data, it is increasingly challenged by modern observations {including late-time cosmology and} early galaxy formation at cosmic dawn (\S \ref{Sec:Intro}) \citep{Bull2016,Peri2022,Efst2025,JWST2024,mvp2024b}.  

{Nevertheless,} the Planck CMB data provide a robust constraint from the BAO on the dimension-full energy density of baryonic and dark matter $\rho_m=\Omega_{m,0}\rho_c$ (\hspace{0cm}\citeauthor{Aghanim2020} \citeyear{Aghanim2020}), where $\rho_c=3H^2/8\pi G$
denotes the closure density given Newton's constant $G$. 
Specifically, Planck provides

\begin{equation}
  c_M=c_M^P\equiv \Omega_{m,0}h^2 = 0.1424\pm 0.00087
  \label{EQN_anchor}
\end{equation}

\noindent where $h=H_0/100$. Planck-$\Lambda$CDM analysis, {whereby,} shows $\Omega_{m,0}=0.3135\pm 0.0073$ with $H_0\simeq 67.36\pm 0.54$ km\,s$^{-1}$\,Mpc$^{-1}$. 

{As mentioned above,} while consistent with the CMB power spectrum, the Planck-$\Lambda$CDM estimate of $H_0$ is inconsistent with LDL, prompting considerations beyond $\Lambda$CDM including a possible dynamical dark energy \citep{mvp2017,Avsa2024,dival2025}. 
In a late-time {spatially three-flat universe}, when radiation can be neglected, {a general formulation of the Hubble evolution satisfies}
(\ref{App_Symmetry})
\noindent

\begin{equation}
    \label{EQN_Hamiltonian}
    H(a)^2=H_0^2\left[\Omega_{m,0}a^{-3}+\left(1-\Omega_{m,0}\right)\hat{\rho}_{DE}\right],
\end{equation}

\noindent where $\hat{\rho}_{DE} = A_0^{-1}A(a)$ is a normalized dark energy, $A(a)=H^2(a)-C_Ma^{-3}$, 
$A_0=\left(H_0^2-C_M\right)=H_0^2\left(1-\Omega_{m,0}\right)$ and $C_M\equiv \Omega_{m,0}H_0^2=(100{\rm \, km\,s^{-1}\,Mpc^{-1}})^2\, c_M$. 
In $w(a)$CDM, $w(a)$ parameterizes $\hat{\rho}_{DE}\equiv\exp(f_{DE})$ based on the continuity equation for dark energy with $p_{DE}=w(a)\rho_{DE}$, 
giving

\begin{equation}
\label{EQN_fde}
 f_{DE}=-3\int_a^1\frac{1+w(a^\prime)}{a^\prime}da^\prime, 
\end{equation}

\noindent Inverting (\ref{EQN_fde}) gives

\begin{equation}
\label{EQN_wcdm}
w(a)=-1-\frac{1}{3}\frac{d\ln \hat{\rho}_{DE}}{d\ln a}
= -1 -\frac{1}{3}\frac{d\ln A(a)}{d\ln a}.
\end{equation}

\noindent {which notably is {\it independent}} of the normalization {constant $A_0$ in $\hat{\rho}_{DE}$}, which drops out upon differentiation. {The resulting $w(a)$ may go through an {a posteriori} linearization \eqref{EQN_w0wa} for ($w_0,\, w_a$)-estimate.}

\textcolor{black}{Given Hubble data $H(a_i)$,
$0<a_i<1$,} (\ref{EQN_wcdm}) shows that $w(a)$ is invariant under perturbations of the matter density that preserve the BAO constraint (\ref{EQN_anchor}), yielding 
(\ref{App_Symmetry})

\begin{eqnarray}
dw(a) \propto dc_M.
\label{EQN_dw}
\end{eqnarray}

\noindent {This invariance makes $w(a)$ insensitive to $\Omega_{m,0}$ or $H_0$, alone, highlighting the role of the BAO constraint \eqref{EQN_anchor}.} 
Accordingly:

\begin{enumerate}
\item $w(a)$ is invariant under {\it constraint-preserving} variations in $\Omega_{m,0}$ obeying \eqref{EQN_anchor};
\item $w(a)$ responds only to {\it constraint-violating} variations in $\Omega_{m,0}$. 
\end{enumerate}

\noindent {Generally, in {departures} from $\Lambda$CDM, \eqref{EQN_anchor} may be subject to changes constraining model parameters.} Including corrections for spectral distortions, DESI values for $c_M$ derived from different background cosmologies remain within about $5\%$ of the Planck value (\ref{EQN_anchor}) \citep{karim2025a}.

In contrast to the exact expression \eqref{EQN_wcdm}, $w_0w_a$CDM, commonly known as CPL, adopts

\begin{equation}
-\left(1+w_0+w_a\right)\ln a + w_a(1-a)
= \frac{1}{3}\ln\left(\frac{A(a)}{A_0}\right).
\label{EQN_HL}
\end{equation}

\noindent Because $A_0$ remains explicit in (\ref{EQN_HL}), the invariance (\ref{EQN_dw}) is lost. Consequently, the model retains a finite sensitivity to constraint-satisfying variations in $\Omega_{m,0}$. This parasitic sensitivity introduces an intrinsic systematic correlation to $\Omega_{m,0}$ even in 
constraint-preserving variations, potentially leading to significant distortions in inferred $w_0w_a$-estimates. Since the loss of (\ref{EQN_dw}) is inherent to early linearization itself, this correlation is generic and independent of the particular data set employed (\ref{App_Symmetry}). in fact, \eqref{EQN_wcdm} and \eqref{EQN_HL} imply that $w_0w_a$ estimates from late linearization are independent of the underlying true $\Omega_{m,0}$, while this invariance is violated for early linearization.

{In light of the above, we  next compare the $w_0w_a$-estimates in (\ref{EQN_w0wa}) obtained from $w(a)$CDM and $w_0w_a$CDM, respectively, corresponding to late and early linearization, the {latter} also applied by DESI. 
We then quantify} the systematic uncertainties introduced by the loss of $c_M$-invariance in
the a priori application of \eqref{EQN_w0wa} through \eqref{EQN_HL} - the early linearization. The resulting correlation between the inferred $w_0w_a$ and the variations in $\Omega_{m,0}$ is demonstrated using controlled mock-data experiments and cosmic chronometer data (\ref{App_data}).

\section{$w_0w_a$ estimates from early and late linearization}
\label{Sec:Data}

{In $w_0w_a$ estimates, we analyze the $H(a)$ data considering $H_0$ and $\Omega_{m,0}$ as control parameters, which may be} anchored by the BAO constraint \eqref{EQN_anchor}. As in DESI, we assume a spatially flat universe described by \eqref{EQN_Hamiltonian} with $\Omega_k=0$.

\begin{figure}[h!]
    \centering
    \includegraphics[width=0.9\linewidth]{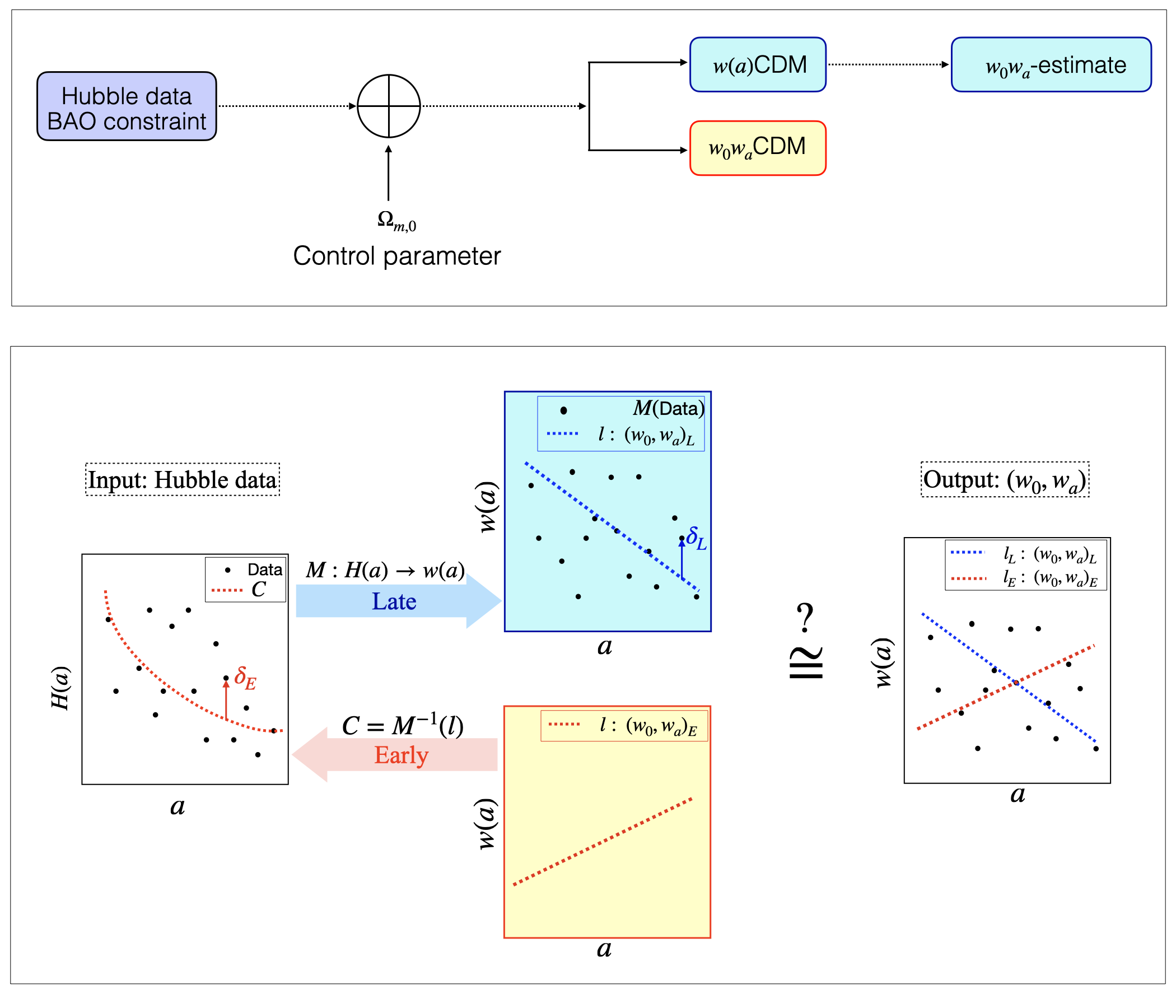}
    \caption{
    ({\it Upper panel}.) 
    Flowchart of late (I) and early (II) linearization according to \eqref{EQN_w0wa} in the $w(a)$CDM {and respectively $w_0w_a$CDM} formalism applied to $H(a_i)$ data with control parameter $\Omega_{m,0}$ (\S\ref{Sec:Data}), optionally {constrained by (\ref{EQN_anchor})}. The two implementations may produce $w_0w_a$-estimates in different quadrants of the $w_0w_a$-plane.
    ({\it Lower panel}.) 
    Schematic of early and late linearization in the estimation of $w_0$ and $w_a$, where $M$ denotes the nonlinear map \eqref{EQN_wcdm}. Early linearization minimizes $\chi^2$ in $H(z)$-space with scatter $\delta_E$, whereas late linearization minimizes $\chi^2$ in $w(z)$-space with scatter $\delta_L$ induced by $M$.
    }.
    \label{fig:implementation}
\end{figure}

Fig.\ref{fig:implementation} {illustrates the approaches we used for $w_0w_a$-estimates} obtained by early and late linearization:

\begin{enumerate}
\renewcommand{\labelenumi}{\textbf{\Roman{enumi}.}}
\item {\it Late linearization:} First extract $w(a_i)$ from \eqref{EQN_wcdm} {using $H(a_i)$ data}, followed by a linear fit using \eqref{EQN_w0wa};
\item {\it Early linearization:} Fit \eqref{EQN_HL} directly {to $H(a_i)$} data in the application of $w_0w_a$CDM.
\end{enumerate}

In late linearization (I), invariance \eqref{EQN_dw} is naturally preserved, whereas in early linearization (II) it is {invariably} violated. Fig.~\ref{fig:implementation} further shows that methods I and II involve different cost functions in the estimation of $w_0w_a$, {i.e. $w(a)$ vs $H(a)$}. Consequently, the resulting estimates generally differ due both to correlations to $\Omega_{m,0}$ intrinsic to method II and to the use of different cost functions. These considerations are generic and independent of the particular data set employed.

To highlight the distinct behavior of these two methods, we quantify the correlation with $\Omega_{m,0}$ inherent in early linearization in $w_0w_a$CDM, considering both constraint-preserving and constraint-violating variations in the control parameter $\Omega_{m,0}\in[0.22,0.36]$ defined by \eqref{EQN_anchor}. This interval is sufficiently broad to cover conventional estimates of $\Omega_{m,0}$ together with $H_0\in[60,80]$ km\,s$^{-1}$\,Mpc$^{-1}$.

\subsection{$\Omega_{m,0}$-correlations in implementation to mock data}
\label{Sec:imp1}

To evaluate the distinct behavior of early and late linearizations, we consider the application of {early and late linearization to mock data produced using CPL according to} \eqref{EQN_HL} and $\Lambda$CDM, over $z \in (0,2)$. We test for consistency in output \textcolor{black}{with control parameters.}
{Without loss of generality,} our mock data are generated with $\Omega_{{m,0}i}=0.27$ satisfying \eqref{EQN_anchor}. For $\Lambda$CDM, $H(a)$ satisfies

\begin{equation}
\label{EQN_LCDM}
H(a)=H_0\sqrt{1+\Omega_{m,0}(a^{-3}-1)}
\end{equation}
with $w_0=-1$ and $w_a=0$.

{To produce mock data using} CPL, we consider the cases $(w_0,w_a)=(-1.5,2)$ and $(w_0,w_a)=(-0.5,-1.5)$ in \eqref{EQN_HL}, representing decreasing and increasing dark-energy densities, respectively. {All three results are then superposed with noise for uncertainties}. Together with the $\Lambda$CDM-generated mock data, this defines an illustrative set of mock $H(a)$ data experiments for early and {late linearizations study, using \eqref{EQN_w0wa},} \eqref{EQN_wcdm}, and \eqref{EQN_HL} (\ref{App_data}).

\begin{figure*}
    \centering
    \includegraphics[width=0.45\linewidth]{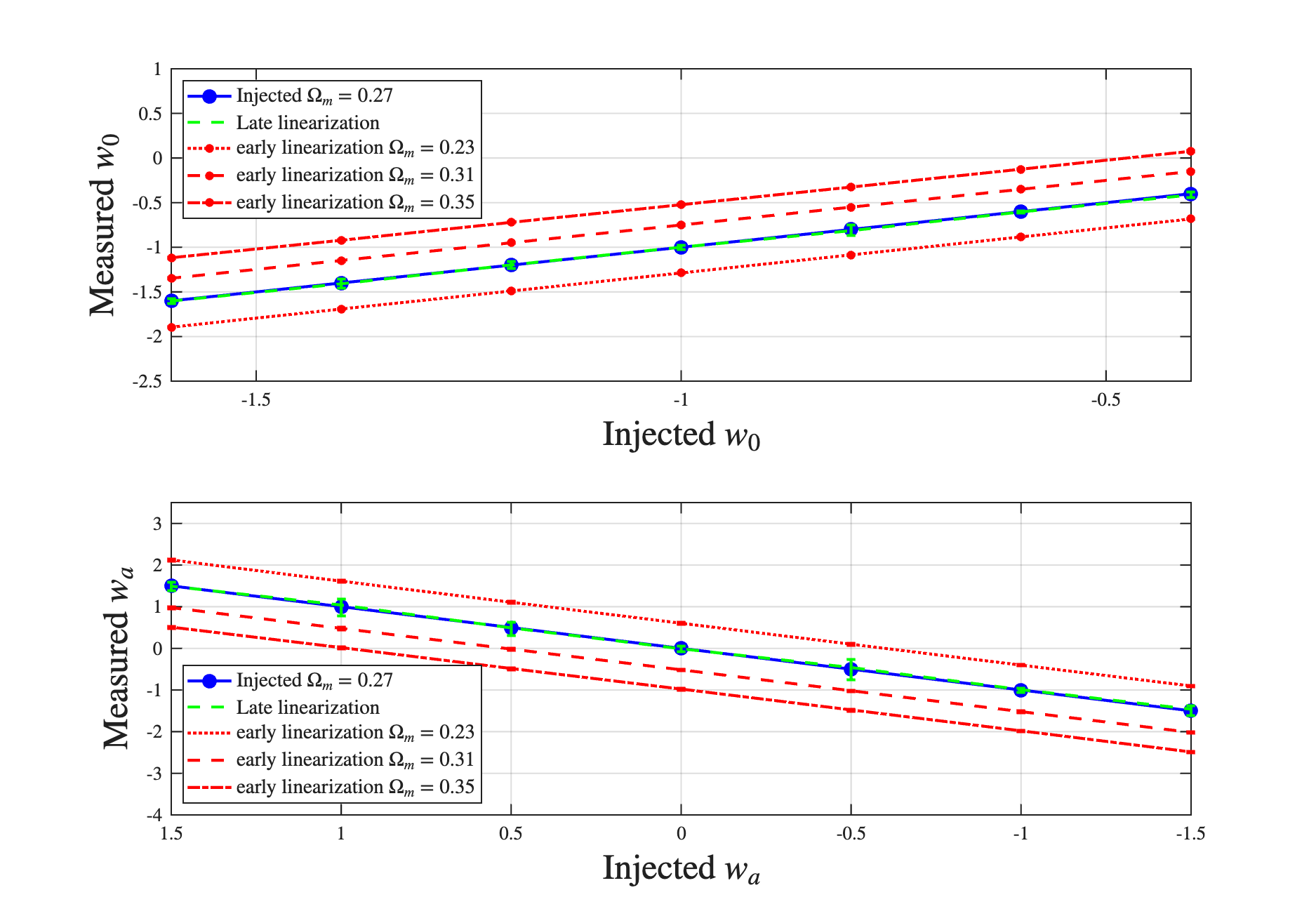}
    \includegraphics[width=0.45\linewidth]{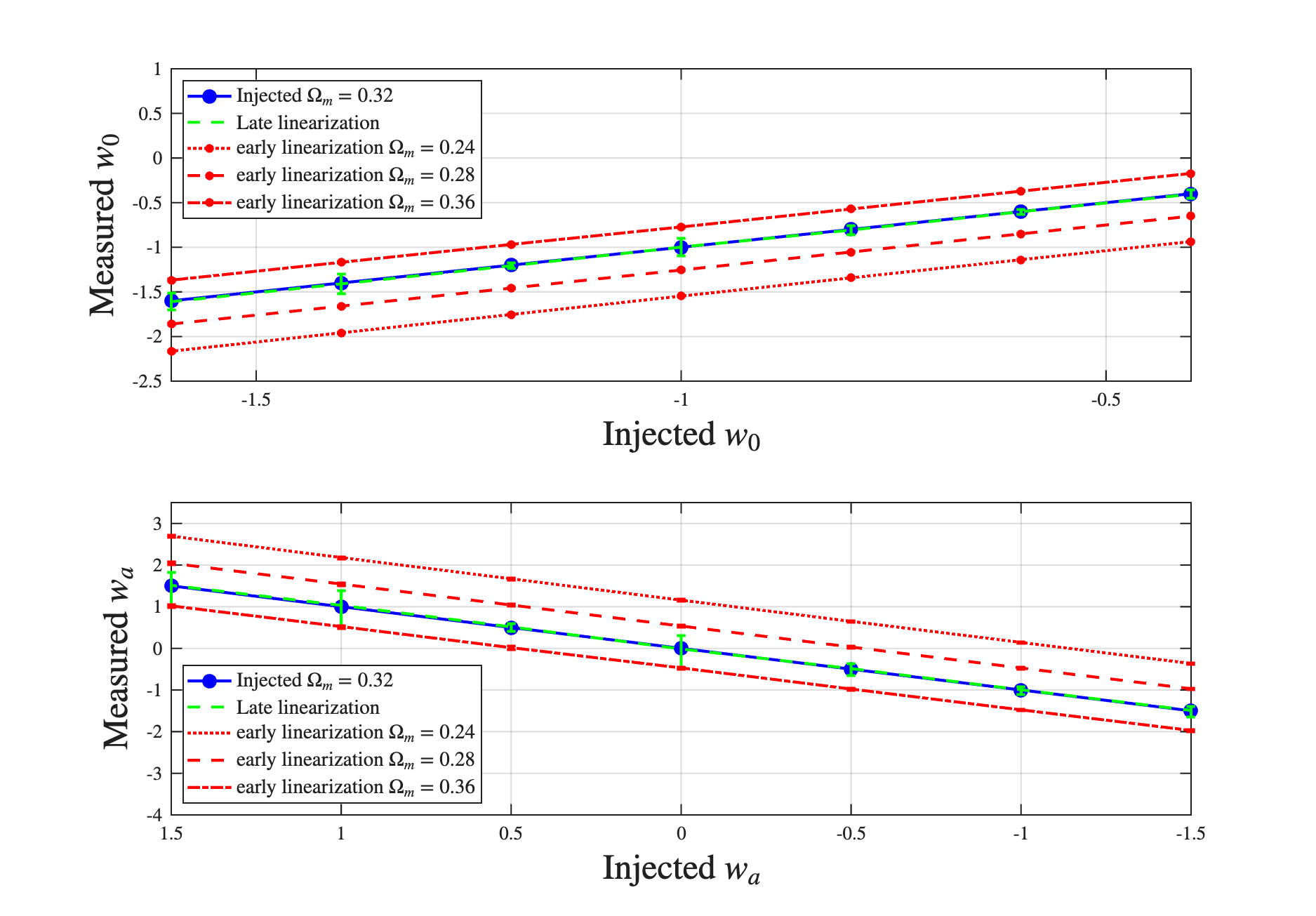}
    \caption{ {Correlations to} $\Omega_{m,0}$ 
    in $w_0w_a$ estimates in mock data experiments on early linearization (red), which is absent in late-linearization (green). 
    Mock data are produced with an injection control parameter $\Omega_{m,0}^{true}$ with the values $\Omega_{{m,0}_i}=0.27$ and, alternatively, $\Omega_{{m,0}_i}=0.32$, each satisfying the BAO constraint (\ref{EQN_anchor}) for a series of $w_0w_a$ injections $[(w_0,w_a)]=[(-1.6,1.5),(-1.4,1),(-1.2,0.5),(-1,0),(-0.8,-0.5),(-0.6,-1),(-0.4,-1.5)]$.
    To emulate mismatches between assumed and true values of $\Omega_{m,0}$, we consider the cases $\Omega_{m,0}=0.23, 0.31, 0.35$ for $\Omega_{{m,0}_i}=0.27$, and $\Omega_{m,0}=0.24, 0.28, 0.36$ for $\Omega_{{m,0}_i}=0.32$, shown in the left and right panels respectively. Note the small errorbars in these figures. While consistent results obtain with late linearization (green) that is insensitive to these mismatches in $\Omega_{m,0}$ as per \eqref{EQN_dw}, inconsistent results obtain with early linearization (red) exhibited by parasitic sensitivity to $\Omega_{m,0}$. 
    In these experiments, $w_0$ from early linearization shows an offset from the true value -- {\it a mismatch} -- 
    representing a sensitivity $S_0=\Delta w_0/\Delta \Omega_{m,0}\simeq 7$; $w_a$ estimates similarly show a sensitivity $S_a=\Delta w_a/\Delta \Omega_{m,0}\simeq -12$. See further Table \ref{tab:3} in \ref{App_sensitivty}).}
    \label{fig:mocksummary}
\end{figure*}

Fig. \ref{fig:mocksummary} reveals inconsistent $w_0w_a$ estimates from {early linearization compared} to initial injection due to its significant sensitivity to constraint-satisfying variations in $\Omega_{m,0}$ under (\ref{EQN_anchor}). In contrast, $w_0w_a$ estimates from late linearization are consistent with the injected values and remain robust against uncertainties in $\Omega_{m,0}$, as implied by \eqref{EQN_wcdm}. The details underlying Fig. \ref{fig:mocksummary} are presented in Fig. \ref{fig:mock}.

To summarize, the inferred $w_0w_a$ by early linearization are related to their true injected values through a {linear correlation} accounting for
$\delta \Omega_{M0} \equiv \Omega_{M0}^{\rm true} - \Omega_{M0}^{\rm inferred}$,

\begin{eqnarray}
\begin{array}{rll}
w_0^{\rm true} &= w_0^{\rm inferred} + S_0 \, \delta \Omega_{M0}, \\\\
w_a^{\rm true} &= w_a^{\rm inferred} + S_a \, \delta \Omega_{M0}
\label{EQN_M0}
\end{array}
\end{eqnarray}

\noindent where $S_i \equiv {\Delta w_i}/{\Delta \Omega_{M0}}$, and $w_i = (w_0, w_a)$ $(i=0,1)$. 
By Fig. \ref{fig:mocksummary}, we conclude

\begin{equation}
S_0 \simeq 7, \,\,S_a\simeq -12.
\label{EQN_SM}
\end{equation}

\noindent Importantly, Fig. \ref{fig:mocksummary} demonstrates that these sensitivities $S_0$ and $S_a$ hold universally for arbitrary choices of $\Omega_{m,0}^{true}$
(see further Table \ref{tab:3} for more details).  

\subsection{$\Omega_{m,0}$-correlation in implementation to cosmic chronometer data}
\label{Sec:imp2}

Applying early and late linearization on clean mock data already proves the inconsistency between early and late linearization.
To further demonstrate the linear correlation (\ref{EQN_M0}-\ref{EQN_SM}), we next turn to cosmic chronometer data tabulated by Farooq et. al. \citep{Faro2017} following \ref{App_data}. These data are particularly illustrative for late-time cosmology \citep{mvp2017a,Abch2021} by sharing a range of redshifts very similar to that of DESI.

Applying the two approaches to cosmic chronometer data, late linearization produces $w_0<-1$ and $w_a>0$ consistent with the results of \eqref{EQN_qw} and with  $w_0$ values derived directly from the data (\ref{App_dataprep}).
In striking contrast, early linearization produces the opposite with $-1<w_0<0$ and $w_a<0$, broadly consistent with DESI results. {Given  \S \ref{Sec:model} and \S \ref{Sec:imp1}}, we attribute this systematic discrepancy with late linearization to the broken invariance of \eqref{EQN_wcdm} under (\ref{EQN_anchor}), arising from the non-commutativity of linearization and the core analysis, {different in $w_0w_a$CDM \eqref{EQN_HL}} and late linearization
posterior to $w(a)$CDM. This discrepancy is possibly compounded by differences in the cost functions of the two. 

\begin{figure}[h!]
\centering
\hspace{0cm}
\includegraphics[width=1\linewidth]{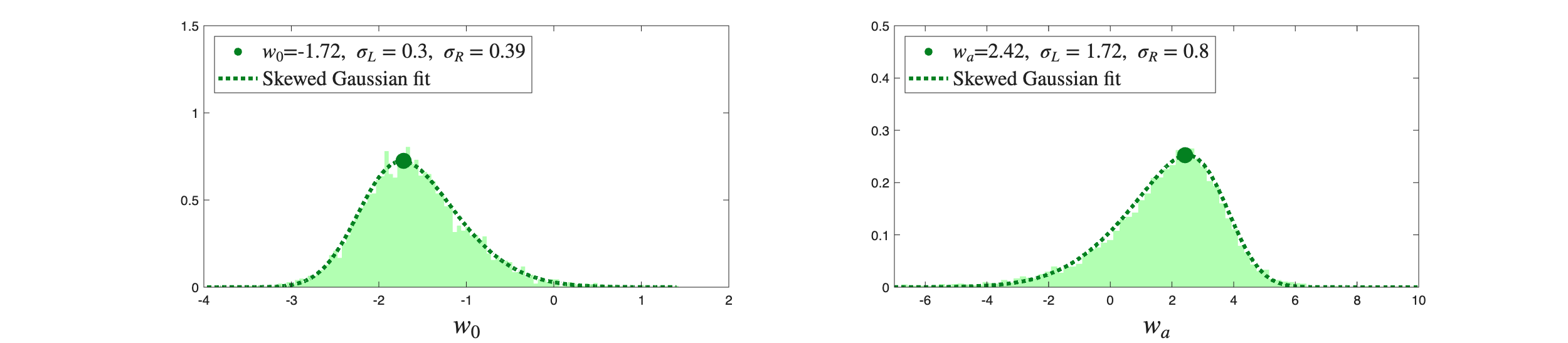}
\hspace{-1cm}
\includegraphics[width=1\linewidth]{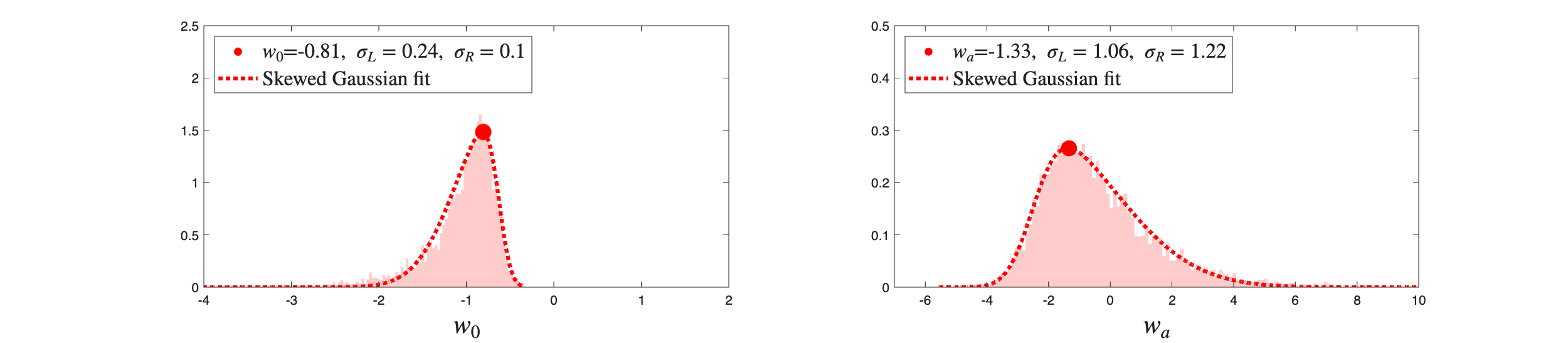}
\caption{\label{fig:histogram} \label{fig:histogramL} ({\it Upper panels.}) 
PDFs of $w_0$ and $w_a$ derived from our MC analysis in late linearization (\S \ref{Sec:Data1}), satisfying invariance with respect to constraint-preserving variations in $\Omega_{m,0}$. 
({\it Lower panels.}) PDFs of $w_0$ and $w_a$ derived from our MC analysis in early linearization (\S \ref{Sec:Data2}), {exhibiting a correlations} to constraint-preserving variations in $\Omega_{m,0}$. 
Here, $\Omega_{m,0}=0.265$ is associated with $H_0=73.3$ km,s$^{-1}$,Mpc$^{-1}$ by \eqref{EQN_anchor}. The peaks of the skewed Gaussian fits define the nominal values of $w_0$ and $w_a$. 
Paradoxically, early linearization (lower panels) reveals a more pronounced nonlinear propagation of uncertainties, {apparent in the appreciably more skewed PDFs} of $(w_0,,w_a)$, than late linearization (upper panels). The resulting discrepancy between these two implementations persists over the entire range $\Omega_{m,0}\in[0.22,0.36]$.}
\end{figure}

Fig. \ref{fig:histogram} and \ref{App_sensitivty} summarize these results, including the implied  sensitivity coefficients $S_0$ and $S_a$ in \eqref{EQN_sensitivity}, which are found to be very similar to those in \eqref{EQN_SM}, {derived from our mock data analysis}.

\section{Interpretation of Results}
\label{Sec:results}

{Prompted by inconsistencies in DESI results for dynamical dark energy with LDL and $\Lambda$CDM, we study possible systematics in probes of dynamical dark energy in the general framework of $w(a)$CDM, arising from  
\textcolor{black}{early} linearization of $w(a)$
\textcolor{black}{in $w_0w_a$CDM.} 
{To this end, we consider \eqref{EQN_w0wa} in late (I) and early (II) linearization in applications to mock and cosmic chronometer data} (\S3 and \ref{App_data}), covering a redshift range similar to that of DESI. Particular attention is paid to the (in)sensitivity {to constraint-satisfying variations} in the matter density $\Omega_{m,0}$ anchored by the BAO constraint (\ref{EQN_anchor}).}

\begin{figure*}[h!]
\centering
\includegraphics[width=0.68\linewidth]{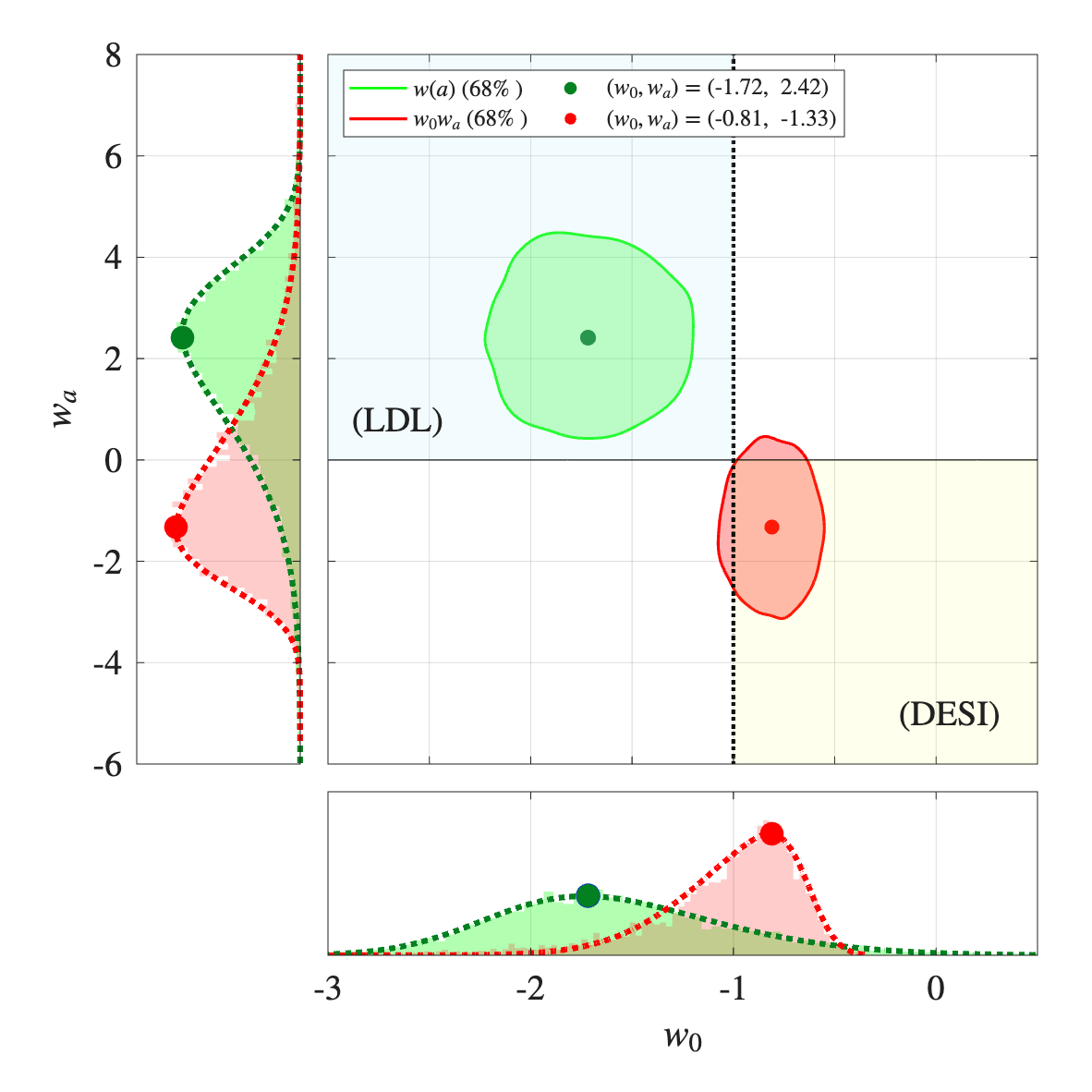}
\caption{\label{fig:Posterior} 
Posterior plot for $w_0$ and $w_a$ in the 
$w_0w_a$ plane in two distinct late and early 
(\ref{Sec:Data1}-\ref{Sec:Data2}) implementations of \eqref{EQN_w0wa}, 
respectively in green and red islands. 
This result strikingly shows a switch in quadrant from IV to II in fallback to the fully nonlinear expression \eqref{EQN_wcdm}, equivalent to a change in the slope and $y$-intercept of the linear approximations in Fig. \ref{fig:TwoLines}. According to \eqref{EQN_dw}, this is due to the violation of an underlying invariance in red, i.e., {\it non-commutativity} between linearization and $w_0w_a$ estimation.
}
\end{figure*}

Fig. \ref{fig:Posterior} and Table \ref{tab:resultsA} summarize the experimental results. 
Due to the correlations with $\Omega_{m,0}$ in early linearization (\ref{EQN_M0}--\ref{EQN_SM}), $w_0w_a$ estimates and the high value of $\Omega_{m,0}$ inferred from this approach are radically different from that obtained by late linearization, which preserves the fully nonlinear expression \eqref{EQN_wcdm} and is free of correlation with $\Omega_{m,0}$. Table \ref{tab:resultsA} shows that the DESI results for these parameters are reproduced in this study using early linearization.

Fig.~\ref{fig:Posterior} shows that the early linearization (\ref{EQN_HL}) places $w_0w_a$ in the fourth quadrant of the $w_0w_a$-plane, effectively {\it antipodal} to its location in the second quadrant obtained by the late linearization. This change in location is evident from the opposite signs of the slopes and the different $y$-intercepts of \eqref{EQN_w0wa} shown in Fig. \ref{fig:TwoLines}.

Importantly, this discrepancy due to the correlation with $\Omega_{m,0}$ persists in our study of mock $H(a)$ data generated by $\Lambda$CDM and CPL (\ref{Sec:imp1}). 
For given injection values of $\Omega_{{m,0}i}$ and $H_0$, Fig. \ref{fig:mock} shows that $w_0w_a$ estimates from late linearization remain consistent with injections and robust against uncertainties in $\Omega_{m,0}$, as implied by \eqref{EQN_wcdm}. Late linearization is thus {robust through} its insensitivity {to constraint-preserving} variations in $\Omega_{m,0}$, {i.e. late linearization is blind to $\Omega_{m,0}$ in estimating $w_0w_a$.} 

By contrast, early linearization does not share this property. The same experiments produce inconsistent $w_0w_a$ estimates whenever $\Omega_{m,0}$ departs from the underlying injected value.

\begin{table*}[h!]
    \centering
    \renewcommand{\arraystretch}{1.5}
    \caption{$w_0,w_a$ estimates from cosmic chronometer data \citep{Faro2017} 
    by early versus late linearization (\S \ref{Sec:Data}) compared with those of Planck $\Lambda$CDM and DESI DR2. {For DESI DR2, $H_0$ is inferred from the DESI estimate for $\Omega_{m,0}$ and the BAO constraint (\ref{EQN_anchor})}.}

    \hspace{-1cm}
    \begin{tabular}{cc|cc||cc|c}
    \hline \hline
        {Parameter} & {Planck $\Lambda$CDM}  & \multicolumn{4}{c|}{{Farooq et.al.\,  data}} & {DESI DR2}\\
         &  &  \multicolumn{2}{c}{Late linearization}& \multicolumn{2}{c|}{Early linearization} &   \\
         \hline \hline
         
        $\Omega_{m,0}$ & $0.3153^{+0.0073}_{-0.0073}$  & 0.265 & 0.352  & 0.265 & \textcolor{teal}{0.352} & \textcolor{teal}{$0.352^{+0.041}_{-0.018}$}\\
        
        $H_0$\,(kms$^{-1}$Mpc$^{-1}$) & $67.36$  &73.30 & 63.60  & 73.30  &\textcolor{teal}{63.60} & \textcolor{teal}{$63.60$}\\
        \hline
        $w_0$ & -1  & $-1.72^{+0.39}_{-0.3}$ & $-1.68^{+0.36}_{-0.34}$ & $-0.81^{+0.1}_{-0.24}$ & \textcolor{teal}{$-0.17^{+0.11}_{-0.25}$} & \textcolor{teal}{$-0.48^{+0.35}_{-0.17}$}\\
        
        $w_a$ & 0  & $2.42^{+0.8}_{-1.72}$  & $2.49^{+0.81}_{-1.76}$ & $-1.33^{+1.22}_{-1.06}$ & \textcolor{teal}{$-2.72^{+1.31}_{-1.11}$} & \textcolor{teal}{$<-1.34$}\\
        \hline
    \end{tabular}
    \label{tab:resultsA}
\end{table*}

Table \ref{tab:resultsA} also lists results for $\Omega_{m,0}=0.265$ with $H_0=73.30$ km\,s$^{-1}$\,Mpc$^{-1}$ consistent with LDL, 
for both methods I and II discussed in \S \ref{Sec:Data}. 
See further \ref{App_sensitivty}  that also presents estimates for constraint-violating variations of $\Omega_{m,0}$, where 
Fig. \ref{fig:6panel} shows {persistent inconsistencies} between early and late linearizations. 

{The sensitivity to $\Omega_{m,0}$ in $w_0w_a$ estimates from early linearization suggests a noticeable correlation in the CPL-based $w_0w_a$ estimates. This points to a potentially serious systematic in CPL-based $w_0w_a$ estimates derived from analyses of stratified data over a finite redshift range, such as those of DESI.}

In contrast, $(w_0,q_0)$ estimates from late linearization are consistent with direct estimates {from data (\ref{App_data})} based on the Taylor expansion $H(z)=H_0(1+(1+q_0)z+\cdots)$. {$(H_0,\Omega_{m,0},w_0,w_a)$ estimates for the cosmic chronometer data are also consistent with LDL.}
Three of these four parameters are independent by virtue of \eqref{EQN_anchor}, while $(w_0,w_a)$ is representative of the deceleration parameter. According to late linearization, $q_0=(3\Omega_{DE,0}w_0+1)/2$ (see \eqref{EQN_qw}), giving $q_0\simeq -1.39$ from the peak value in Fig. \ref{fig:histogram}. This outcome is consistent with earlier estimates of $q_0$ in the LDL framework \citep{mvp2017a,Cama2020,Simo2025}.
 
Our $w_0$ and $w_a$ {estimates depart from the DESI estimates by $2\sigma$ and $1.6\sigma$}, respectively. 
Furthermore, imposing \eqref{EQN_anchor} with $H_0=73.04$ km\,s$^{-1}$\,Mpc$^{-1}$ from LDL \citep{Ries2022} {consistent with that of cosmic chronometer data}, gives $\Omega_{m,0}=0.268$, which departs by about $4.66\sigma$ from the DESI DR2 estimate. If indeed the true $\Omega_{m,0}$ is similarly low, the combined discrepancy in $\Omega_{m,0}$ and $w_0w_a$ estimates suggests an overall tension of about $4.94\sigma$ with DESI. 

Fig. \ref{fig:Posterior} illustrates this discrepancy through the location of the green island (late linearization in II), consistent with LDL, and the red island (early linearization in IV), representative of DESI. Roughly equivalent in terms of their central values, $q_{0,g}\simeq -1.39$ for green and $q_{0,r}\simeq -0.4$ for red, we obtain
\noindent
\begin{eqnarray}
q_{0,g}<q_0^{\Lambda}<q_{0,r}.
\end{eqnarray}

\noindent The first inequality is consistent with previous LDL estimates of $q_0$ and indicates a dark energy density that is increasing, in contrast to the second inequality, which implies the opposite behavior.

In all cases considered in our analysis extended over the range of $\Omega_{m,0}$ as control parameter, 
the red island is found to be significantly unstable with respect to variations in $\Omega_{m,0}$ 
due to {the $\Omega_{m,0}$ dependency} inherent to early linearization (\ref{App_sensitivty}). 

{This result persists} independently of the data, observed in analyses applied to different data sets, here mock data, 
cosmic chronometer data of \cite{Faro2017} and {\cite{Moresco2022}; {see further \ref{Sec:imp1} and \ref{App_dataprep}.}}
Moreover, we find the same resulting from non-commutativity between parametrization and the fully non-linear $w(a)$CDM analysis, 
to persist in non-linear approaches \citep{lodha2025} beyond linearization \eqref{EQN_w0wa}.

\section{Conclusion}
\label{Sec:conclusion}

In this work we consider cosmological parameter estimation to probe for a dynamical dark energy in $w(a)$CDM {defined} by the equation of state $w(a)$ {and its linearization methods} (\S \ref{Sec:model}) in light of results by DESI, LDL and Planck-$\Lambda$CDM. Specifically, we consider anchoring in the Planck BAO constraint (\ref{EQN_anchor}).

{We do so pursuant to late and early linearization applied to mock data and cosmic chronometer data {(\S \ref{Sec:Data} and \S \ref{App_dataprep})}. The results of early and late linearization, {\it ceteris paribus}, are remarkably distinct due to loss of the invariance \eqref{EQN_dw} under constraint-preserving variations in $\Omega_{m,0}$ in the first, which is otherwise preserved in the second.} Our probe is expressed in conventional $w_0w_a$ estimates (\ref{EQN_w0wa}).

 We test for systematics in early linearization — the CPL approach — expected from $\Omega_{m,0}$-correlations (\ref{EQN_M0}-\ref{EQN_SM}), otherwise absent in the fully nonlinear formulation $w(a)$CDM (\ref{EQN_dw}).
The distinct results from early and late linearization due to the mismatch between $\Omega_{{m,0}_i}$ and $\Omega_{m,0}$ is demonstrated using clean mock data (Fig. \ref{fig:mocksummary}).

Fig. \ref{fig:Posterior} shows the resulting locations in the $w_0w_a$ {plane when the approaches are applied to cosmic chronometer data}: 

\begin{itemize}
\item Green islands in quadrant II for late linearization following the nonlinear map \eqref{EQN_wcdm} and are free of $\Omega_{m,0}$-correlations by \eqref{EQN_dw}. Its location indicates an increasing dark energy consistent with LDL and in agreement with $q_0$ and $w_0$ derived directly from $H(z)$ (Tables \ref{table:Hzdata} and \ref{tab:resultsA}), including results from Pantheon data over $z<0.2$ \citep{Cama2020}, confirmed by Pantheon and Pantheon+ data over the full range $z\lesssim2.26$ \citep{Simo2025}.

\item Red islands in quadrant IV for early linearization, in contrast, indicate a decreasing {(thawing)} dark energy based on (\ref{EQN_HL}). {This result is consistent with DESI} (Table \ref{tab:resultsA}), but is likely in $4.94\sigma$ disagreement with $q_0$ and $w_0$ directly derived from LDL data.
\end{itemize}

The distinct locations of the red and green islands in the $w_0w_a$ plane persist regardless of the choice of data (\ref{Sec:Data}) due to persistent $\Omega_{m,0}$-correlations in early linearization whenever the inferred $\Omega_{m,0}$ does not equal the underlying true value (Fig. \ref{fig:mocksummary}). This is potentially relevant for DESI, which reports an inferred matter density $\Omega_{m,0}\simeq 0.35$ that appears significantly larger than the LDL and Planck values, satisfying $0.26\lesssim \Omega_{m,0}\lesssim 0.32$.

We recall that CPL is historically derived from $w(a)$CDM, providing the basis for our comparison in the application of the two approaches to the data, for which Eq. \eqref{EQN_M0} represents an $\Omega_{m,0}$-bias. However, one may also consider these two approaches side-by-side, as if posited independently. In this interpretation, Eq. \eqref{EQN_M0} represents an interaction between (dark) matter density and dark energy, otherwise absent in the initial $w(a)$CDM wherein the two are non-interacting.

{{\it If} the true $\Omega_{m,0}$ is close to the LDL-anchored value and the non-interacting $w(a)$CDM framework holds, the relations (10)-(11) would map the DESI estimate $w_0 = -0.48^{+0.35}_{-0.17}$ (Table \ref{tab:resultsA}) to $-1.2 < w_0 < -0.7$. This mapping assumes that (11), evaluated with fixed values of the control parameter $\Omega_{m,0}$, carries over to joint-interference pipelines such as DESI's, a detailed demonstration of which falls outside the scope of the present analysis.
Under the alternative face-value reading of CPL, no such revision of the DESI $w_0$-value applies.}
Although frequently stated \cite[e.g.][]{cal03,cli04,lud17}, we note that the possibility $w_0 < -1$ does not necessarily violate the null energy condition (NEC) or imply ghost instabilities. See, for instance, \cite{mvp2025,van26} for a geometric framework in which this regime is natural, exemplified by various data analyses in \cite{Moresco2022}.

Combined with similar bias corrections for $w_a$, this raises the possibility of dark energy residing in quadrant II of the $w_0w_a$ plane if the true matter density satisfies $\Omega_{m,0}\simeq 0.27$.

Independent confirmation of the nature of dark energy may come from the current {\it Euclid} survey of about $\mathcal{O}(10^9)$ galaxies up to a redshift of a few (\hspace{0cm}\citeauthor{Euclid}\, \citeyear{Euclid}). 

\section*{Acknowledgment}
The authors thank the anonymous reviewer for constructive comments that greatly improved the clarity of the presentation. This work is supported by the National Research Foundation of Korea, under Grants No. NRF-RS-2024-00334550. 

\bibliographystyle{elsarticle-harv} 
\bibliography{DESI-bib}{}

\clearpage
\appendix
\section{The underlying invariance}
\label{App_Symmetry}
In $w(a){\rm CDM}$, {dark energy is defined} by the equation of state $w(a)$ that satisfies the energy conservation law and the continuity equation. In a spatially flat universe at late times, the Hamiltonian energy constraint is
\begin{equation}
\label{EQN_threeflat}
\Omega_m+\Omega_{DE}=1,
\end{equation}
where $\Omega_m=\rho_m/\rho_c$ and $\Omega_{DE}=\rho_{DE}/\rho_c$ are the dimensionless matter and dark energy densities, normalized to the closure density $\rho_c=3H^2/8\pi G$, with $H\equiv H(a)$ the Hubble parameter.

Cold baryonic and dark matter densities scale with $a^{-3}$, i.e., $\rho_m=\rho_{m,0}a^{-3}$ and $\Omega_{m,0}=\rho_{m,0}/\rho_{c,0}$, where $\rho_c=\rho_{c,0}E^2(a)$, $E(a)=H(a)/H_0$ and $H_0$ is the Hubble constant.
Normalizing $\rho_{DE}$ to $\rho_{c,0}$,
$\hat{\rho}_{DE}\equiv \rho_{DE}/\rho_{c,0}=\Omega_{DE}E^2(a)$, \eqref{EQN_threeflat} and \eqref{EQN_Hamiltonian} give
\begin{eqnarray}
\hat{\rho}_{DE} = \frac{A(a)}{A_0}\equiv
\frac{H(a)^2-C_Ma^{-3}}{\left(1-\Omega_{m,0}\right)H_0^2},
\label{EQN_A}
\end{eqnarray}
where 
$C_M\equiv \Omega_{m,0}H_0^2=(100{\rm \, km\,s^{-1}\,Mpc^{-1}})^2\, c_M$.

 {In the $w(a)$CDM formalism, $w(a)$ is defined by} $\hat{\rho}_{DE}\equiv e^{f_{DE}}$ in terms of $f_{DE}$ in \eqref{EQN_fde}, based on the continuity equation for {dark energy taking} (\ref{EQN_Hamiltonian}) into
$H^2(a)=H_0^2\left[\Omega_{m,0}a^{-3}+(1-\Omega_{m,0})e^{f_{DE}}\right]$.
Accordingly, we arrive at \eqref{EQN_wcdm}, 
{\it independent of the normalization constant $A_0$}, which drops out upon differentiation. 
\textcolor{black}{This contains a subtle highlight: data input is strictly $H(a_i)$ without $H_0$.}

The BAO in the power spectrum of the CMB provides a primary cosmological constraint \eqref{EQN_anchor}, whose Planck-$\Lambda$CDM \eqref{EQN_anchor} is generally considered to be precise. 
But it turns out to be slightly model dependent with DESI value to be within 5\% of the Planck-$\Lambda$CDM one.
{Data $(a_i,h(a_i)=H(a_i)/H_0)$ define $A(a)=H(a)^2-C_Ma^{-3}$ over $0<a_i<1$.} 
{In this regard,} (\ref{EQN_wcdm}) implies

\begin{eqnarray}
dw = - \left(\frac{2}{3} \frac{aHH^\prime}{A}+1+\frac{C_Ma^{-3}}{A}\right)\left(\frac{a^{-3}}{A}\right) dC_M \propto dc_M,
\label{EQN_dwA}
\end{eqnarray}

\noindent {which defines the invariance of $w(a)$ under constraint-satisfying variations} $\delta c_M = h^2\delta\Omega_{m,0}+\Omega_{m,0}\delta h^2=0$, {leaving sensitivity only to constraint-violating variations $\delta c_M\ne 0$.}
Following Fig. \ref{fig:implementation}, this invariance carries over to $w_0w_a$ estimates by late linearization based on $\chi^2$ fits in the $w(a)$ plane. {we consider this invariance in two distinct linearization approaches discussed in \S \ref{Sec:model}.}

In contrast with the {invariance-preserving approach} above, early linearization considers direct projection onto $w_0w_a$ — the null hypothesis of this analysis — allowing integration of $f_{DE}$ up front, giving the simplified relation \eqref{EQN_HL}. In this approximation, data $(a_i,h(a_i))$ together with control parameters $\Omega_{m,0}$ and $H_0$ allow for $w_0w_a$ estimates from $\chi^2$ fits to (\ref{EQN_HL}) in the $(a,h(a))$ plane (Fig. \ref{fig:mocksummary}). This tests {the null-hypothesis of a linear $w(a)$} relation with or without imposing the constraint (\ref{EQN_anchor}).

The explicit appearance of $A_0$ in (\ref{EQN_HL}), otherwise absent in (\ref{EQN_wcdm}), {introduces dependency} on $H_0$ that breaks the invariance (\ref{EQN_dw}) of late linearization of the same general formulation, $w(a)$CDM. In $w_0w_a$ estimates, therefore:

\begin{enumerate}
\item Late linearization preserves invariance with respect to constraint-preserving variations in $\Omega_{m,0}$;
\item Early linearization breaks this invariance through explicit sensitivity to $H_0$.
\end{enumerate}

\noindent This distinction gives rise to an $\Omega_{m,0}$-correlation in early linearization, regardless of the choice of $c_M$, with significant observational consequences for probing the nature of dark energy in the late-time Universe.

\section{Analysis of Cosmic Chronometer Hubble data}
\label{App_data}

Cosmic chronometer data tabulated by Farooq et. al. \citep{Faro2017} trace the late-time Universe over a redshift range roughly similar to that of DESI.
They are listed as a function of redshift $z_i>0$ in Table \ref{table:Hzdata}, which may be converted to $0<a_i=1/(1+z_i)<1$ for the purpose of our analyses.

\begin{table}[h]
\begin{center}
\scriptsize
\begin{threeparttable}
\caption{Hubble parameter versus redshift data}
\begin{tabular}{cccc}
\hline\hline
~~$z_i$ & ~~$H(z)$(km s$^{-1}$ Mpc $^{-1}$) &~~~~~~ $\sigma_{H}$(km s$^{-1}$ Mpc $^{-1}$) &~~~ Reference\\
\hline\\*[-4pt]
0.070&~~	69&~~~~~~~	19.6&~~~ 5\\

0.090&~~	69&~~~~~~~	12&~~~ 1\\

0.120&~~	68.6&~~~~~~	26.2&~~~ 5\\

0.170&~~	83&~~~~~~~	8&~~~ 1\\

0.179&~~	75&~~~~~~~	4&~~~ 3\\

0.199&~~	75&~~~~~~~	5&~~~ 3\\

0.200&~~	72.9&~~~~~~	29.6&~~~ 5\\

0.270&~~	77&~~~~~~~	14&~~~ 1\\

0.280&~~	88.8&~~~~~~	36.6&~~~ 5\\

0.352&~~	83&~~~~~~~	14&~~~ 3\\

0.380&~~	81.5&~~~~~	1.9&~~~ 10\\

0.3802&~~	83&~~~~~	13.5&~~~ 9\\

0.400&~~	95&~~~~~~~	17&~~~ 1\\

0.4004&~~	77&~~~~~	10.2&~~~ 9\\

0.4247&~~	87.1&~~~~~	11.2&~~~ 9\\

0.440$^*$&~~	82.6&~~~~~	7.8&~~~ 4\\

0.4497&~~	92.8&~~~~~	12.9&~~~ 9\\

0.4783&~~	80.9&~~~~~	9&~~~ 9\\

0.480&~~	97&~~~~~~~	62&~~~ 2\\

0.510$^*$&~~	90.4&~~~~~	1.9&~~~ 10\\

0.593&~~	104&~~~~~~	13&~~~ 3\\

0.600$^*$&~~	87.9&~~~~~	6.1&~~~ 4\\

0.610$^*$&~~	97.3&~~~~~	2.1&~~~ 10\\

0.680&~~	92&~~~~~~~	8&~~~ 3\\

0.730$^*$&~~	97.3&~~~~~	7&~~~ 4\\

0.781&~~	105&~~~~~~	12&~~~ 3\\

0.875&~~	125&~~~~~~	17&~~~ 3\\

0.880&~~	90&~~~~~~~	40&~~~ 2\\

0.900&~~	117&~~~~~~	23&~~~ 1\\

1.037&~~	154&~~~~~~	20&~~~ 3\\	

1.300&~~	168&~~~~~~	17&~~~ 1\\

1.363&~~	160&~~~~~~	33.6&~~~ 8\\

1.430&~~	177&~~~~~~	18&~~~ 1\\

1.530&~~	140&~~~~~~~	14&~~~ 1\\

1.750&~~	202&~~~~~~~	40&~~~ 1\\

1.965&~~	186.5&~~~~~	50.4&~~~ 8\\

2.340$^*$&~~	222&~~~~~~~	7&~~~ 7\\

2.360$^*$&~~	226&~~~~~~~	8&~~~ 6 \\ 

\hline\hline

\label{table:Hzdata}
\end{tabular}
\begin{tablenotes}
\item -Reprinted from \citep{Faro2017}.
\item * {These data have underlying $\Lambda$CDM dependencies and are excluded from our analysis.}
\item { 
(1) \cite{Simon2005}; 
(2) \cite{Stern2010}; 
(3) \cite{Moresco2012}; 
(4) \cite{Blak2012}; 
(5) \cite{Zhang2014}; 
(6) \cite{FontRibera2014}; 
(7) \cite{Delubac2015}; 
(8) \cite{Moresco2015}; 
(9) \cite{Moresco2016}; 
(10) \cite{Alam2017}.}
\end{tablenotes}
\end{threeparttable}
\end{center}
\end{table}

These data are processed according to Fig. \ref{fig:implementation} to study late (I) versus 
early (II) linearization.
In I, $\chi^2$ is implemented in $w(z)$-space according to scatter $\delta_L$ in $w(a)$ space in Fig. \ref{fig:implementation}. 
In II, $\chi^2$ is implemented in $H(z)$ space 
according to the scatter $\delta_E$.

\subsection{Data preparation}
\label{App_dataprep}

\begin{figure}
    \centering
    \includegraphics[width=1\linewidth]{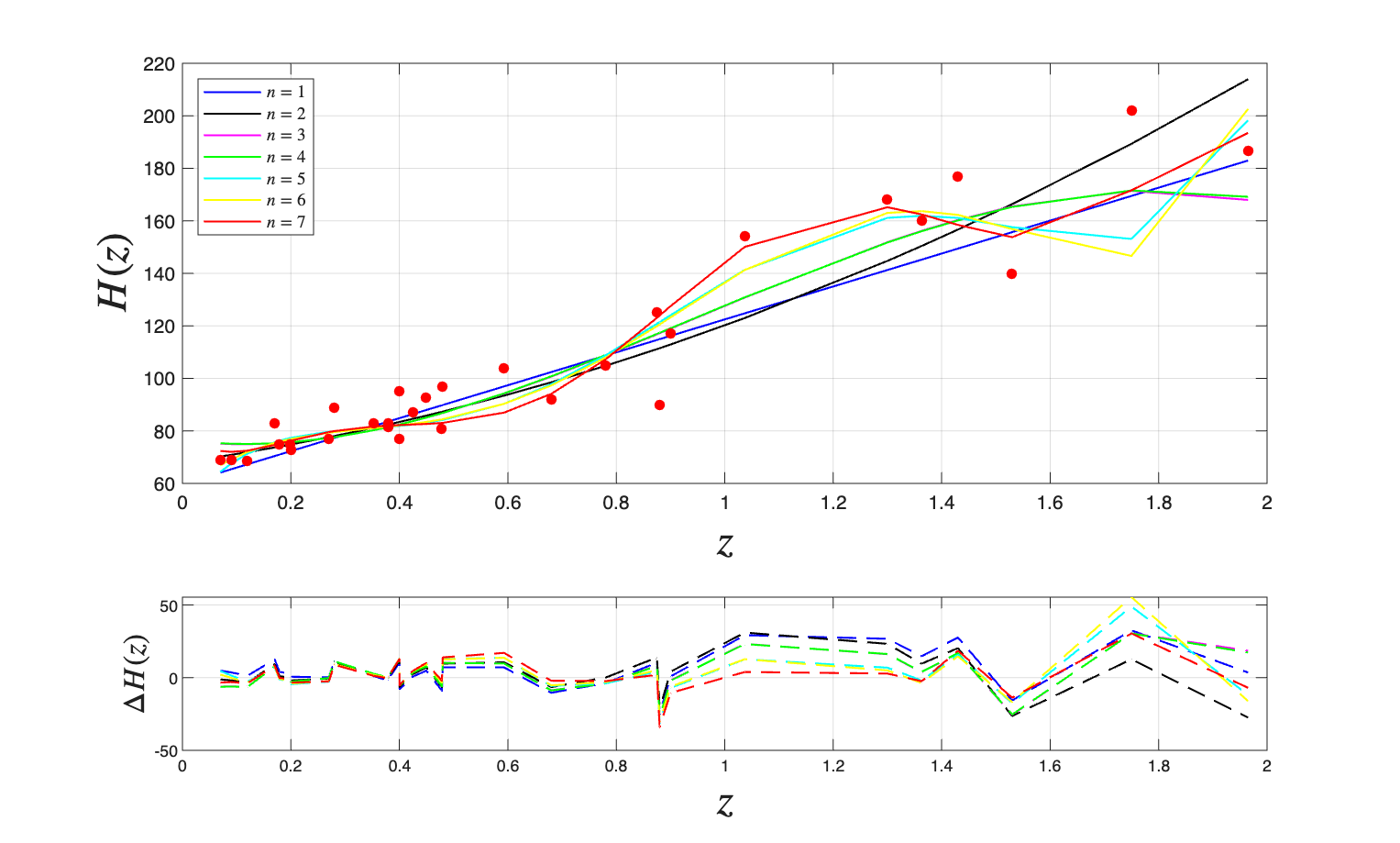}
    \caption{Goodness of fit for polynomials of orders $n=1,2,3,4,5,6,7$ to cosmic chronometer data and their residuals, showing the expected reduction in residuals with increasing $n$.}
    \label{fig:DifferentOrders}
\end{figure}

Starting from the {cosmic chronometer} data in Table \ref{table:Hzdata} (or the mock data in \ref{Sec:Data}), we first perform a smoothing procedure that slightly tightens the standard deviations in the output using a {global polynomial fit of order 3} with weights $1/\sigma_i^2$ \citep{mvp2017, Abch2021}. We have confirmed that this smoothing is free of systematics by varying the polynomial degree and comparing the output with unsmoothed data. The results of this comparison is summarized in Fig. \ref{fig:DifferentOrders}. 

{In the analysis of the cosmic chronometer data of \cite{moresco2020,Moresco2022}, the covariance matrix is strongly diagonally dominated, as indicated by the small normalized off-diagonal elements. Mock-data tests equivalent to zero off-diagonal covariance yield consistent parameter sensitivities, supporting the neglect of correlations in the likelihood.}

To estimate $w_0w_a$ including uncertainties, we apply an MC process with $N=5000$ trials defined by randomly selected Hubble parameter data within the interval $H(z_i)\pm1\sigma_i$ of the Farooq data. Each trial is smoothed by the aforementioned cubic fit. This process also generates a PDF of $H_0$ and $q_0$, derived here from the first two terms in the Taylor expansion of $H(z)$ about $z=0$. For mock data, consistent parameter values are recovered from the control parameter $\Omega_{m,0}$.

For cosmic chronometer data, the resulting estimates $H_0=75.88_{-4.59}^{+4.79}$ km\,s$^{-1}$\,Mpc$^{-1}$ and $q_0=-1.39^{+0.33}_{-0.28}$ are consistent with LDL. In particular, the $q_0$ estimate is consistent with $q_0=-1.08\pm0.29$ derived independently from the Pantheon sample of type Ia  Supernovae \citep{Cama2020,Simo2025}. The combined estimate shows a $3\sigma$ departure from $q_0^{\Lambda}$ of the 
Planck-$\Lambda$CDM estimate, 
$q_0^{\Lambda}\simeq -0.527$. 
For a {spatially flat} universe where radiation can be neglected

\begin{equation}
    w(z)=\frac{2q(z)-1}{3\Omega_{DE}(z)},
    \label{EQN_qw}
\end{equation}

\noindent {giving $w_0=w(z=0)=-1.8_{-0.27}^{+0.32}$, below} the $\Lambda$CDM value $w_0=-1$ (see also \cite{Simo2025}). Largely model-independent, these $q_0$ and $w_0$ estimates, {derived directly from the data,} serve as reference values to the two implementations (I) and (II).

In what follows we discuss the implementation of early and late linearization for cosmic chronometer data in some more detail than our analysis of mock data. 

\subsection{Late linearization}
\label{Sec:Data1}

We realize late linearization in implementation (I) with a large-$N$ MC process. {For each trial it produces} $w(a_i)$ values over all $a_i=1/(1+z_i)$ of Table \ref{table:Hzdata} using the invertible map \eqref{EQN_wcdm}. Implementation (I) thereby moves the analysis from $H(z)$ to $w(a)$ space (Fig. \ref{fig:implementation}) with no loss of information. In this approach, $H_0$ derives from \eqref{EQN_anchor} following a choice of the control parameter $\Omega_{m,0}$. For $H_0$ from the LDL, this corresponds to $\Omega_{m,0}\simeq0.265$.

We next extract $w_0w_a$ estimates through a linear fit to $w(z_i)$ using \eqref{EQN_w0wa} (Fig. \ref{fig:histogram}). According to \ref{Sec:model}, these $w_0w_a$ estimates are invariant under constraint-preserving variations in $\Omega_{m,0}$.

\subsection{Early linearization}
\label{Sec:Data2}
{Next, we turn to the second implementation II, defined by {\it a priori} inclusion of linearization \eqref{EQN_w0wa},
in $H(z)$ space defined in \eqref{EQN_HL}.}
Fig. \ref{fig:histogramL} shows the resulting PDFs of $w_0w_a$ estimates for a choice of $\Omega_{m,0}$. 

\begin{figure}[h!]
\centering
\includegraphics[width=0.95\linewidth]{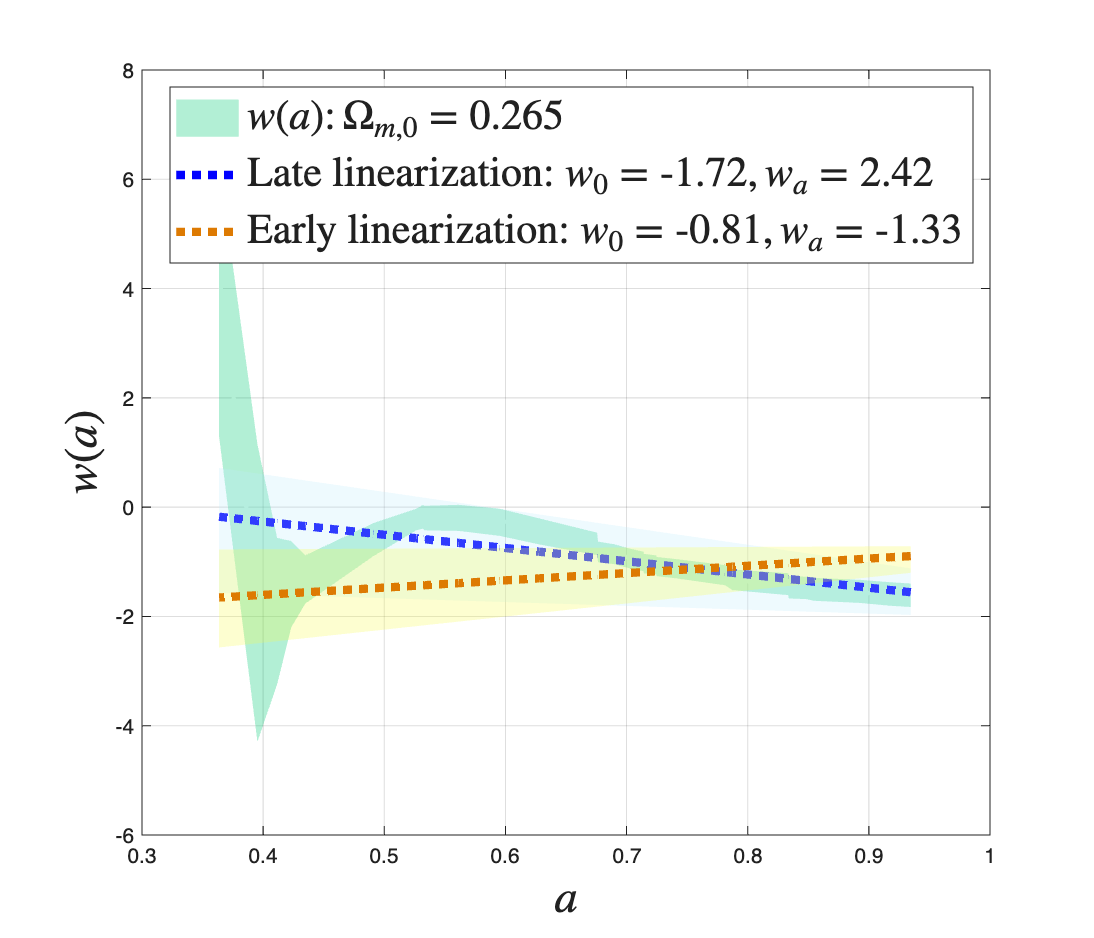}
\caption{\label{fig:TwoLines} 
The equation of state $w(a)$ in \eqref{EQN_wcdm} 
and the $w_0w_a$ estimate \eqref{EQN_w0wa} by slope and $y$-intercept in late and early implementations following Fig. \ref{fig:implementation}.
The green shaded region from late linearization shows $w(a_i)$ transformed under the map \eqref{EQN_wcdm} including an MC process. 
The blue dotted line by \eqref{EQN_w0wa} applied to $w(a_i)$ (\S \ref{Sec:Data1}) reveals a negative slope indicative of $w_a>0$. 
The red dotted line from early linearization, on the other hand, results by application of \eqref{EQN_HL} suggests a positive slope. 
These distinct slopes demonstrate non-commutativity of $w_0w_a$ estimation and linearization of $w(a)$CDM highlighted in Fig. \ref{fig:implementation}.}
\end{figure}

Fig. \ref{fig:TwoLines} shows the resulting  
$w_0w_a$ estimates for the two approaches I and II, showing contradictory outputs for dark energy behavior.
That is, $w_0w_a$ estimation and linearization 
of $w(a)$CDM does not commute.

\begin{figure}
    \centering
\includegraphics[width=1\linewidth]{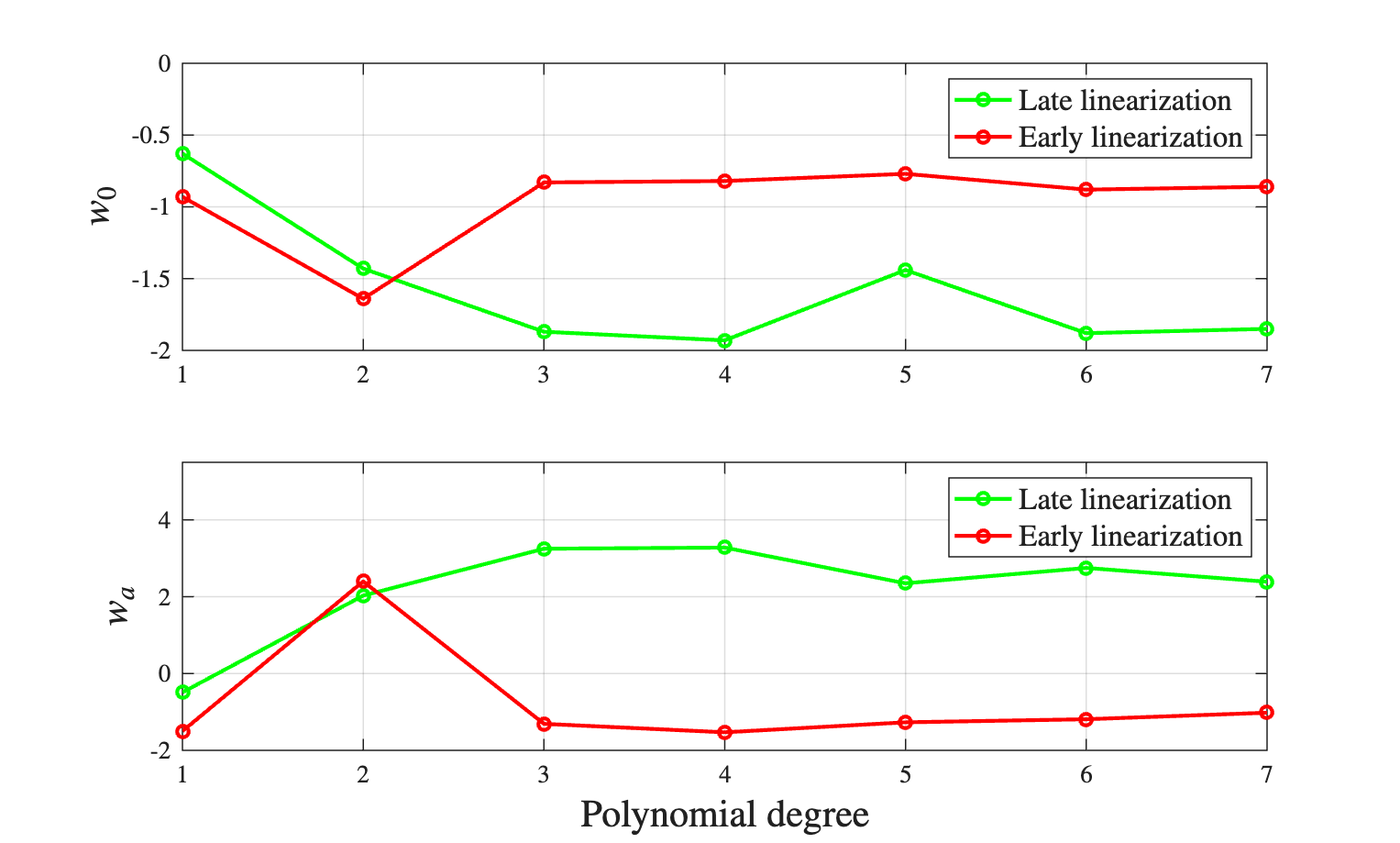}
    \caption{$w_0w_a$ estimates using global polynomial smoothing functions of order $n=1. . 7$. The estimated values are stable with respect to the polynomial order when $n\geq 3$. Here $\Omega_{m,0}=0.265$ associated with $H_0=73.3$ km\,s$^{-1}$\,Mpc$^{-1}$ by \eqref{EQN_anchor}.}
    \label{fig:DifferentOrders2}
\end{figure}

Using a higher order global polynomial for smoothing for early and late linearization, we show $w_0w_a$ estimates are stable for $n\geq 3$ (Fig. \ref{fig:DifferentOrders2}).  
We confirm robustness hereof in \S \ref{Sec:Data1} and \S \ref{Sec:Data2}, further by using locally weighted polynomial regression of order three, summarized in Table \ref{tab:3}.

The discrepancy between late and early linearization, here shown for CPL, persists also for other parametrizations of  \cite{lodha2025}. This highlights persistence of $\Omega_{m,0}$-correlations whenever parametrizations are done a priori, regardless of the model at hand.

\section{Analysis of mock data}
\label{App_mock}
\begin{figure*}[h]
     \centering
    \includegraphics[width=0.75\linewidth]{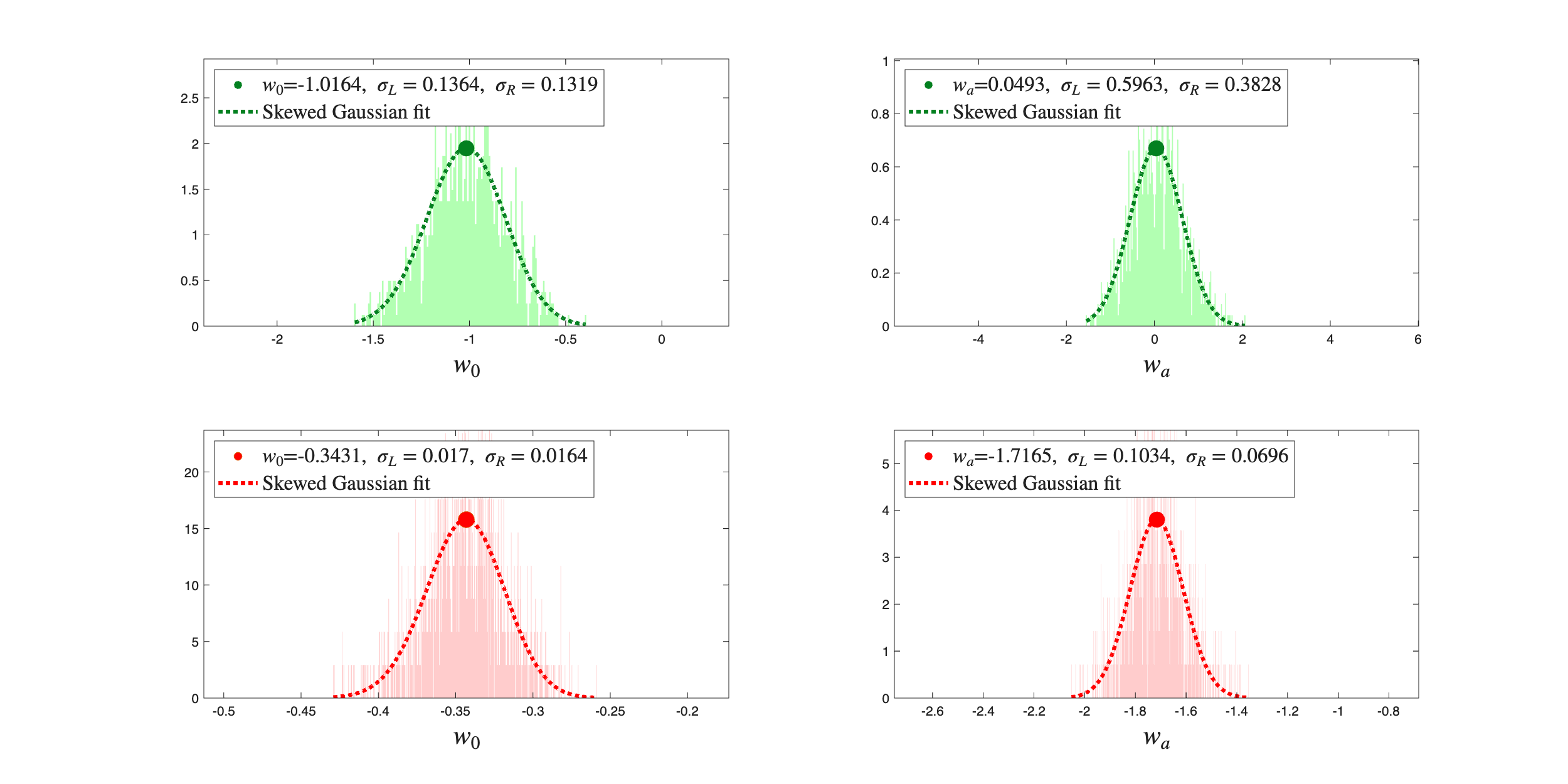}
    \includegraphics[width=0.75\linewidth]{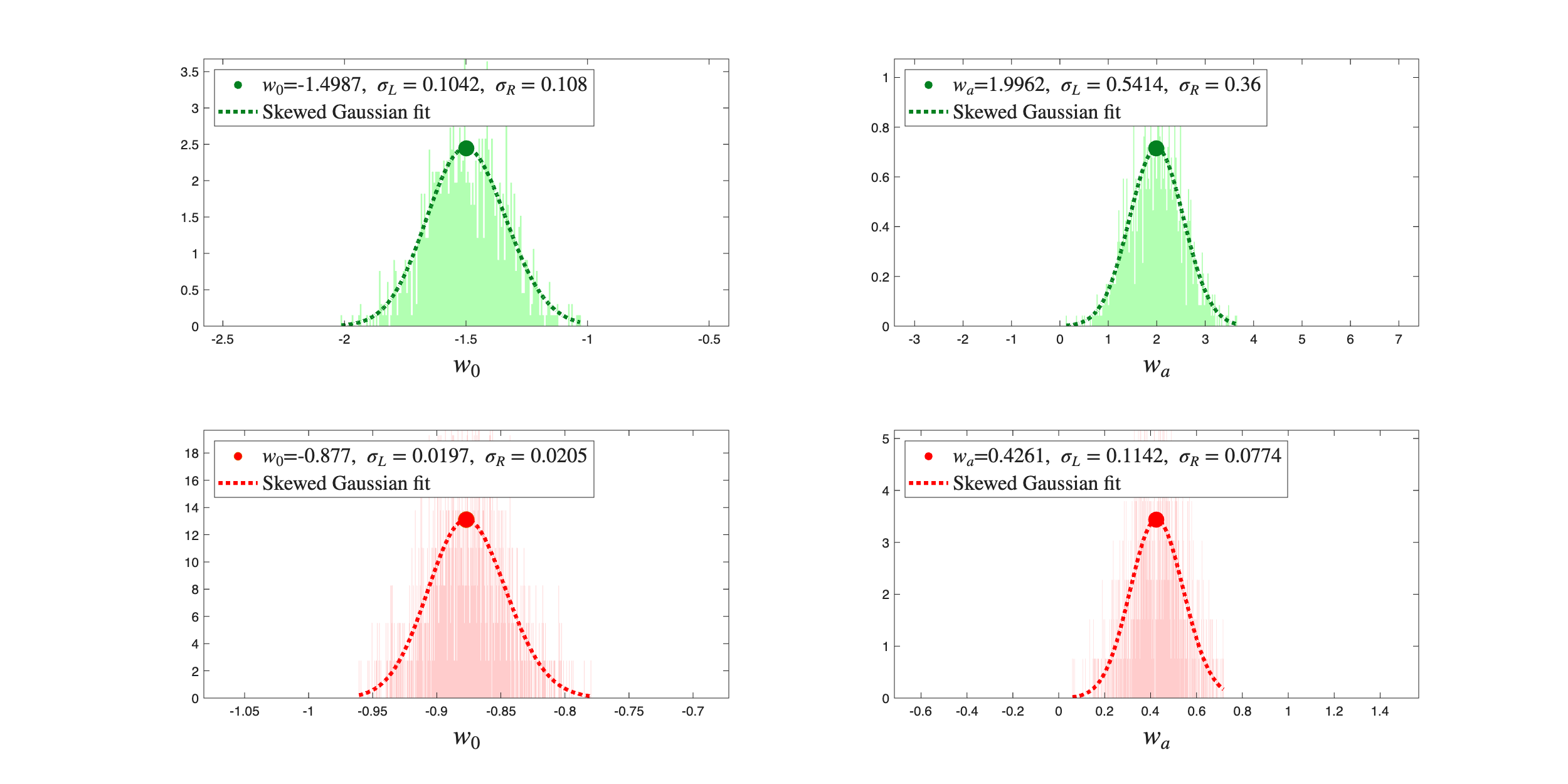}
    \includegraphics[width=0.75\linewidth]{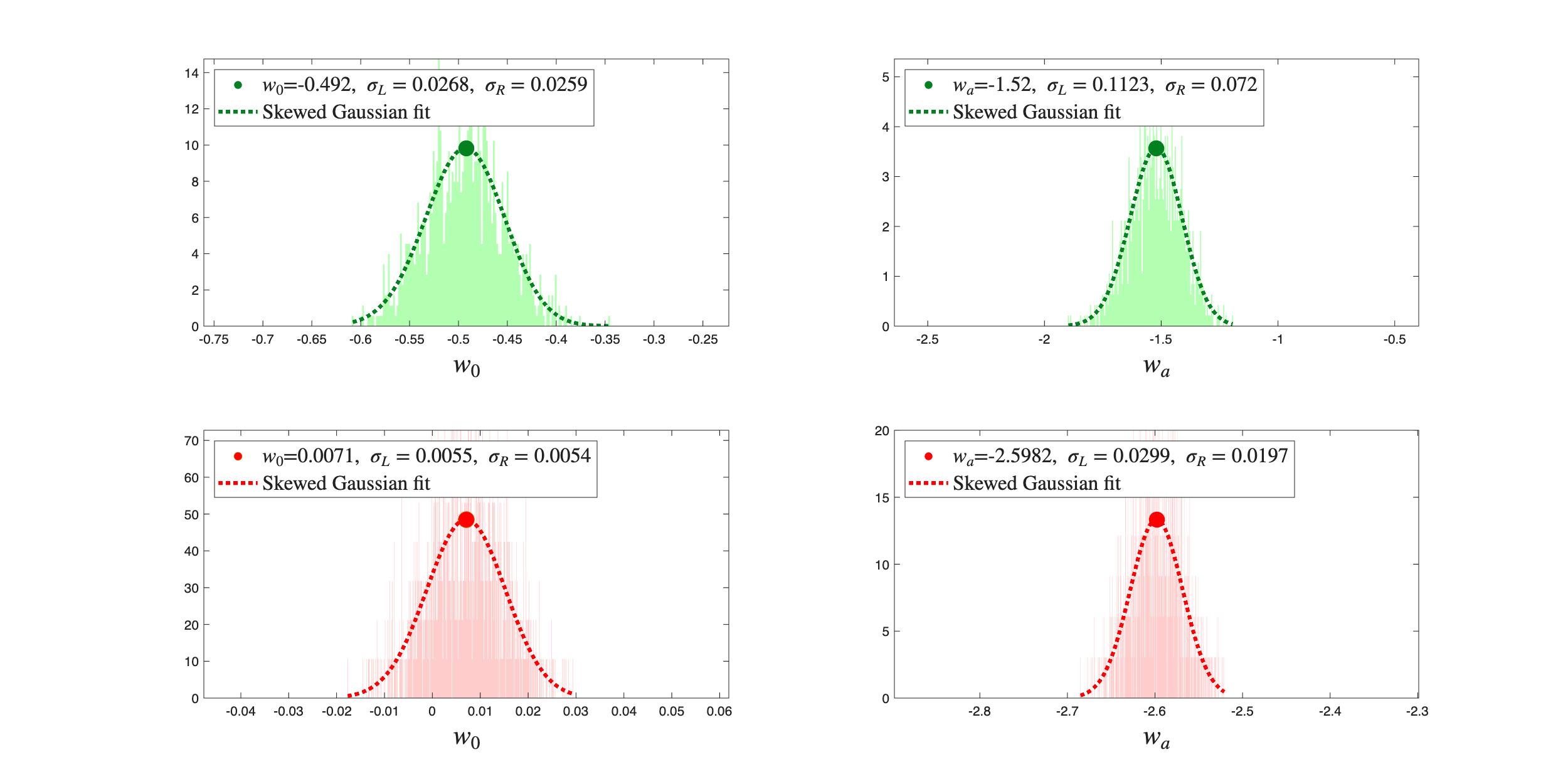}
    \captionof{figure}
    {$w_0w_a$ estimates by early and late linearization. ({\it Four upper panels}). 
    $w_0w_a$ estimate on $\Lambda$CDM mock data using the injection value $\Omega_{{m,0}_i}=0.27$ with $w_0=-1$ and $w_a=0$. Shown is a snapshot for the case of assuming $\Omega_{m,0}=0.35$ in the $w_0w_a$ estimation process. 
    Green (late linearization) gives consistent results with injection values regardless of $\Omega_{m,0}$, as per (5). Red (early linearization), in contrast, exhibits parasitic sensitivity to the control parameter by $\Omega_{m,0}$-correlation (see also \eqref{EQN_sensitivity}), which is consistent with green {\it only} when assumed equals the underlying true value: $\Omega_{m,0} = \Omega_{{m,0}_{i}}$    
({\it Four middle panels}). 
   Idem for data produced by CPL with injection values $w_0=-1.5$ and $w_a=2$. Green gives results consistent with the injections, while red fails.
({\it Four lower panels}). 
Idem using $w_0=-0.5$ and $w_a=-1.5$ with the same discrepant results between the two methods. These examples demonstrate the non-commutativity of $w_0w_a$ estimation and linearization of $w(a)$CDM, manifested through $\Omega_{m,0}$-correlations.}
    \label{fig:mock}
\end{figure*}

For reference, we include a mock-data analysis including $\Lambda$CDM, again for both linearizations using injections $(w_0,w_a)=(-1,0)$, $(w_0,w_a)=(-1.5,2)$, and $(w_0,w_a)=(-0.5,-1.5)$. Fig. \ref{fig:mock} shows the PDFs for the resulting $w_0w_a$ estimates {in a large-$N$ MC process}. Here, we select $\Omega_{{m,0}i}=0.27$ for injection, while $w_0w_a$ estimation {assumes $\Omega_m\in[0.22,0.35]$} — {taking snapshot samples herein for testing $\Omega_{m,0}$-correlations in each method.}

Fig. \ref{fig:mock} shows the output obtained when assuming $\Omega_{{m,0}}=0.35$. {As expected from} (\ref{Sec:model}), late linearization accurately follows the injection, while early linearization reveals its $\Omega_{m,0}$-correlations when $\Omega_{m,0}\neq\Omega_{{m,0}i}$. This corresponds to a $+30\%$ deviation from the injected $\Omega{{m,0}_i}$, which results in $\Delta w_0\sim+0.55$ due to broken invariance in \eqref{EQN_dw}.

\section{Sensitivity to $\Omega_{m,0}$ and $c_M$}
\label{App_sensitivty}
Fig. \ref{fig:Posterior} shows $w_0w_a$ estimates 
in the analysis of cosmic chronometer data, revealing significant $\Omega_{m,0}$-correlations in early linearization.

To illustrate this further, we also study the sensitivity to the BAO constraint (\ref{EQN_anchor}), by varying $c_M$ with $\pm 20\%$ about the Planck value.
As before, we use the injection value $\Omega_{m,0}=0.265$ of LDL and assume the DESI value $\Omega_{m,0}=0.352$ in the $w_0w_a$ estimation. 

\begin{table*}[h!]
\centering
\caption{{Sensitivity to $\Omega_{m,0}$ in the analyses of various data sets using different types of smoothing.} }

\begin{tabular}{|C{3cm}|C{1.8cm}|C{1.8cm}||C{1.8cm}|C{1.8cm}||C{1.8cm}|C{1.8cm}|}

\hline

& \multicolumn{2}{c||}{No Smoothing}
& \multicolumn{2}{c||}{Globally weighted polynomial}
& \multicolumn{2}{c|}{Locally weighted polynomial} \\

\cline{2-7}

Data
& $S_0$ & $S_a$
& $S_0$ & $S_a$
& $S_0$ & $S_a$ \\

\hline

\hspace{0.5cm}Mock data \textcolor{white}{Mock datadata}
& $6.5^{+0.002}_{-0.01}$
& $-13.5^{+0.007}_{-0.01}$
& $5.7^{+0.01}_{-0.01}$
& $-11.7^{+0.003}_{-0.05}$
& $5.8^{+0.02}_{-0.02}$
& $-12.2^{+0.05}_{-0.08}$ \\

\hline

Cosmic Chronometer (\cite{Faro2017})
& $4.9^{+0.14}_{-0.18}$
& $-15.6^{+0.95}_{-0.93}$
& $7.1^{+0.13}_{-0.12}$
& $-18^{+0.8}_{-0.9}$
& $7.7^{+0.12}_{-0.1}$
& $-16.5^{+0.73}_{-0.84}$ \\

\hline

Cosmic Chronometer (\cite{Moresco2022})
& $6.9^{+0.2}_{-0.15}$
& $-10.4^{+1}_{-0.9}$
& $7.5^{+0.15}_{-0.13}$
& $-13.3^{+0.85}_{-0.8}$
& $6.5^{+0.15}_{-0.14}$
& $-15^{+0.72}_{-0.84}$ \\

\hline

\end{tabular}

\label{tab:3}
\end{table*}

\begin{figure*}[h!]
    \centering
    \includegraphics[width=0.48\linewidth]{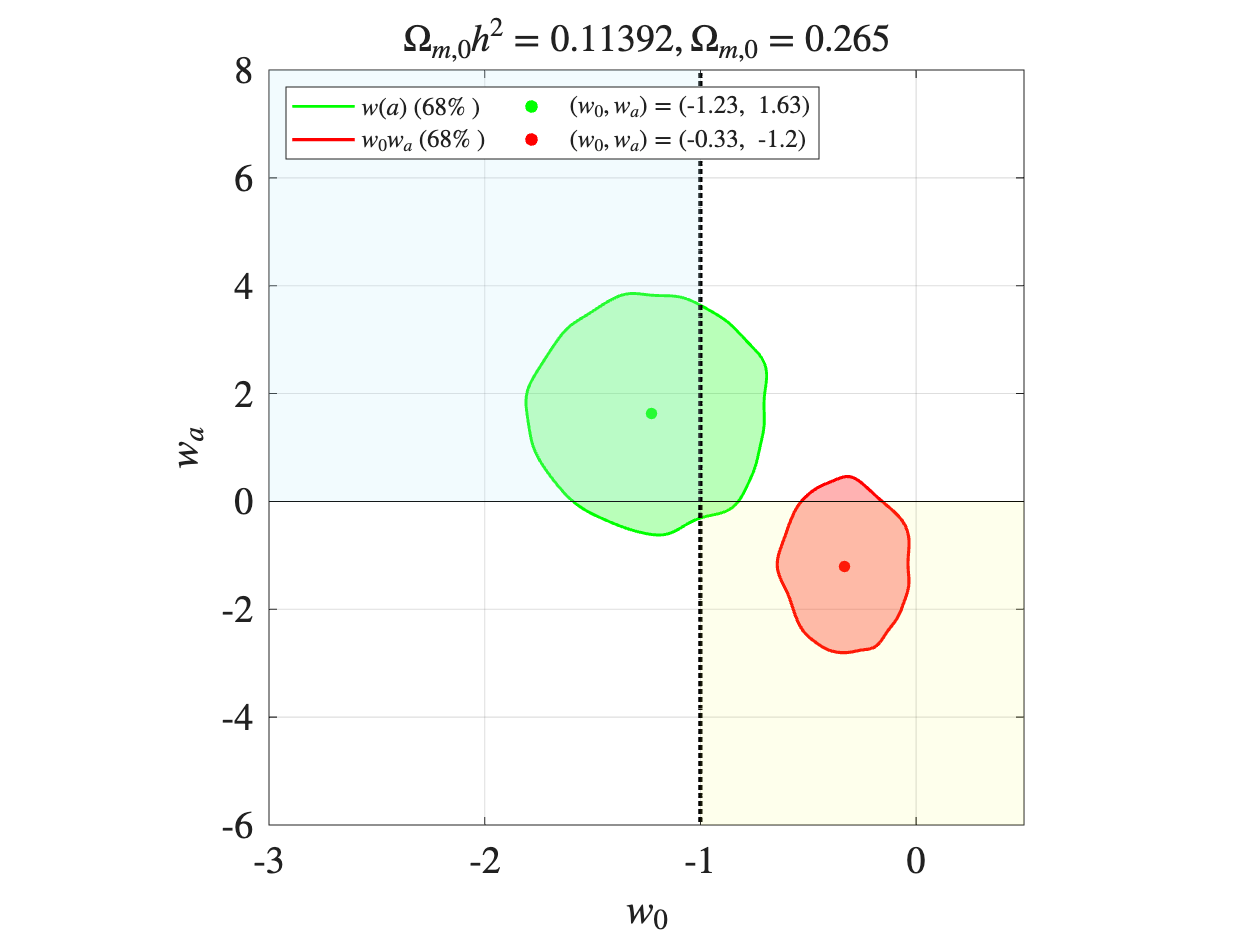}
    \includegraphics[width=0.48\linewidth]{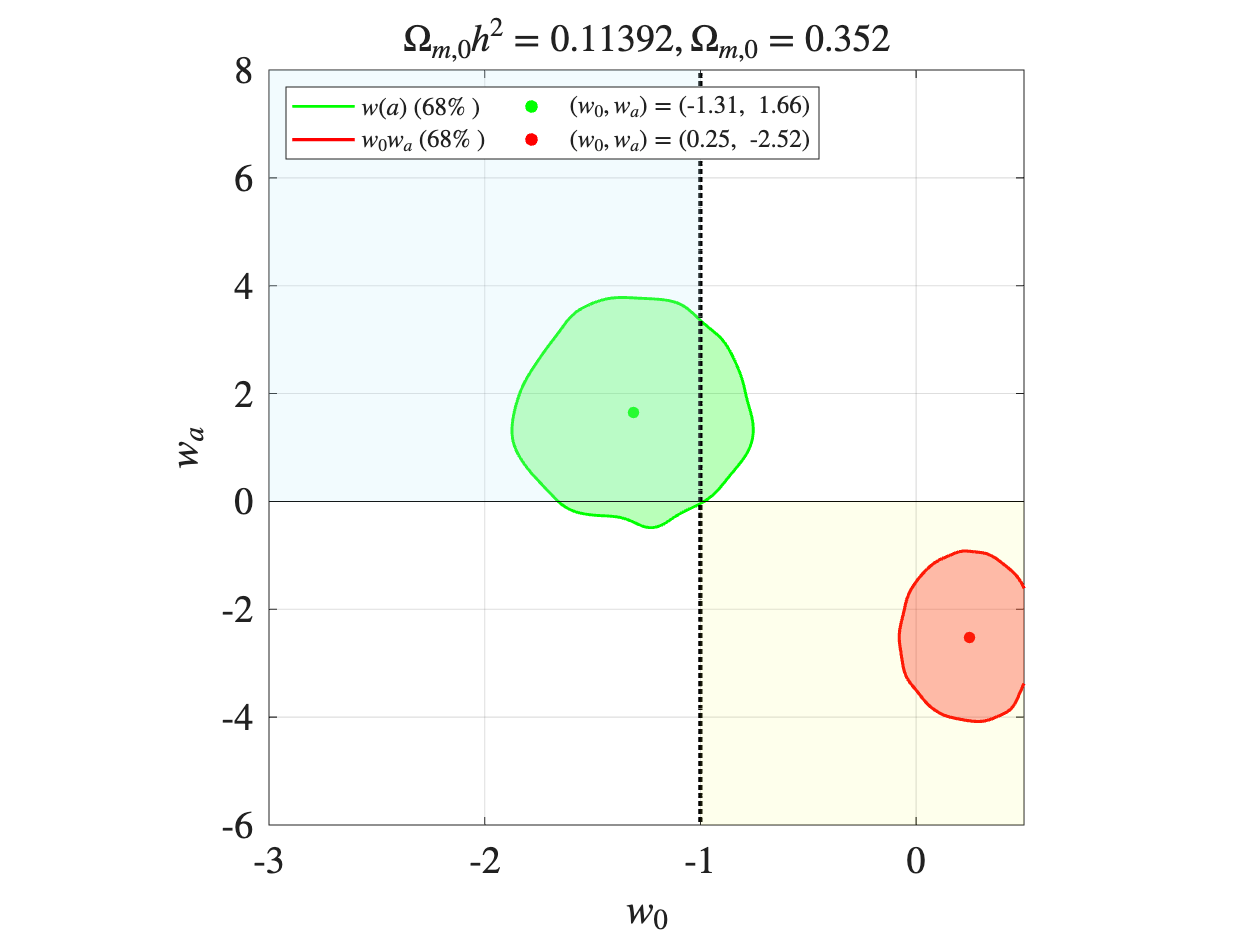}
    \includegraphics[width=0.48\linewidth]{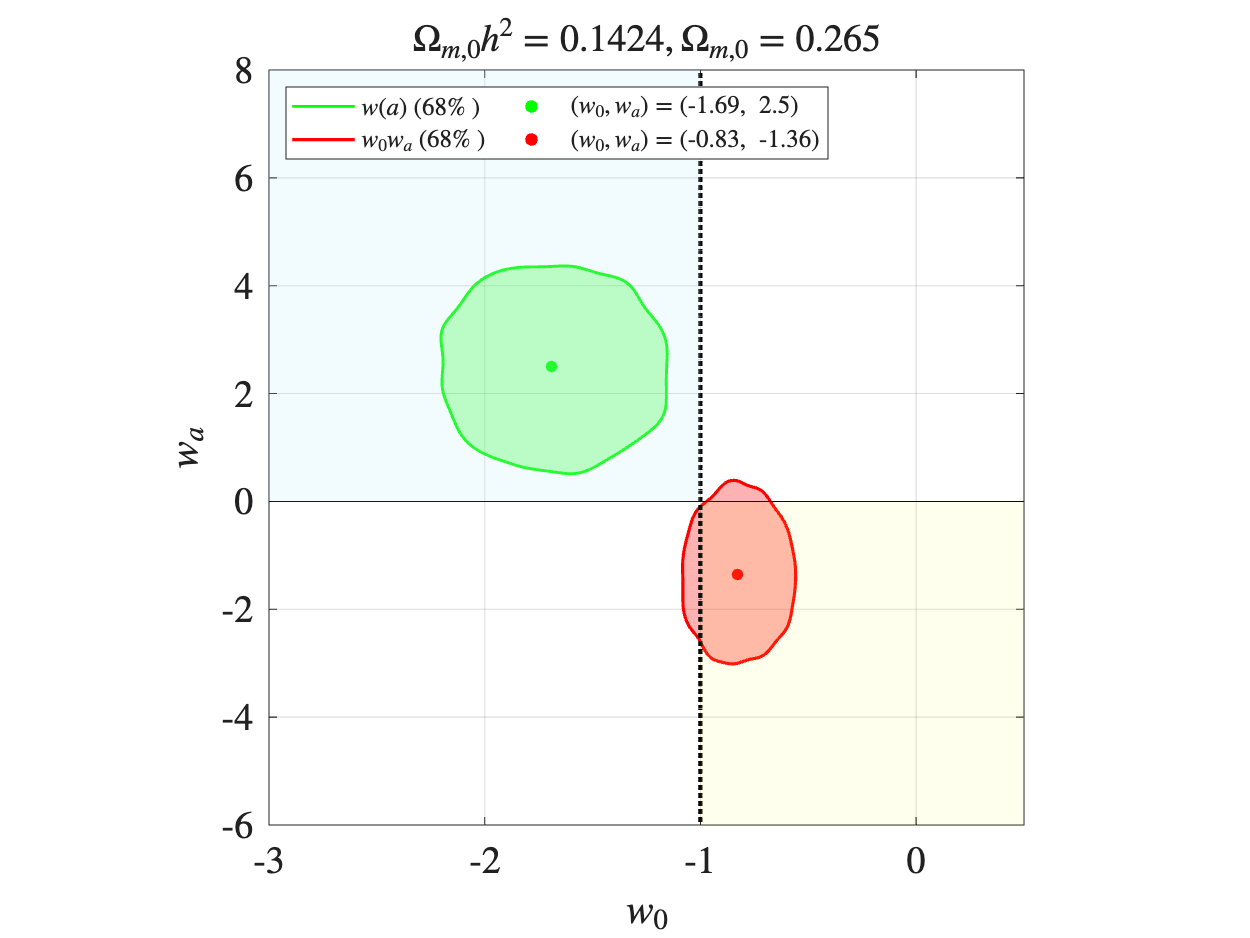}
    \includegraphics[width=0.48\linewidth]{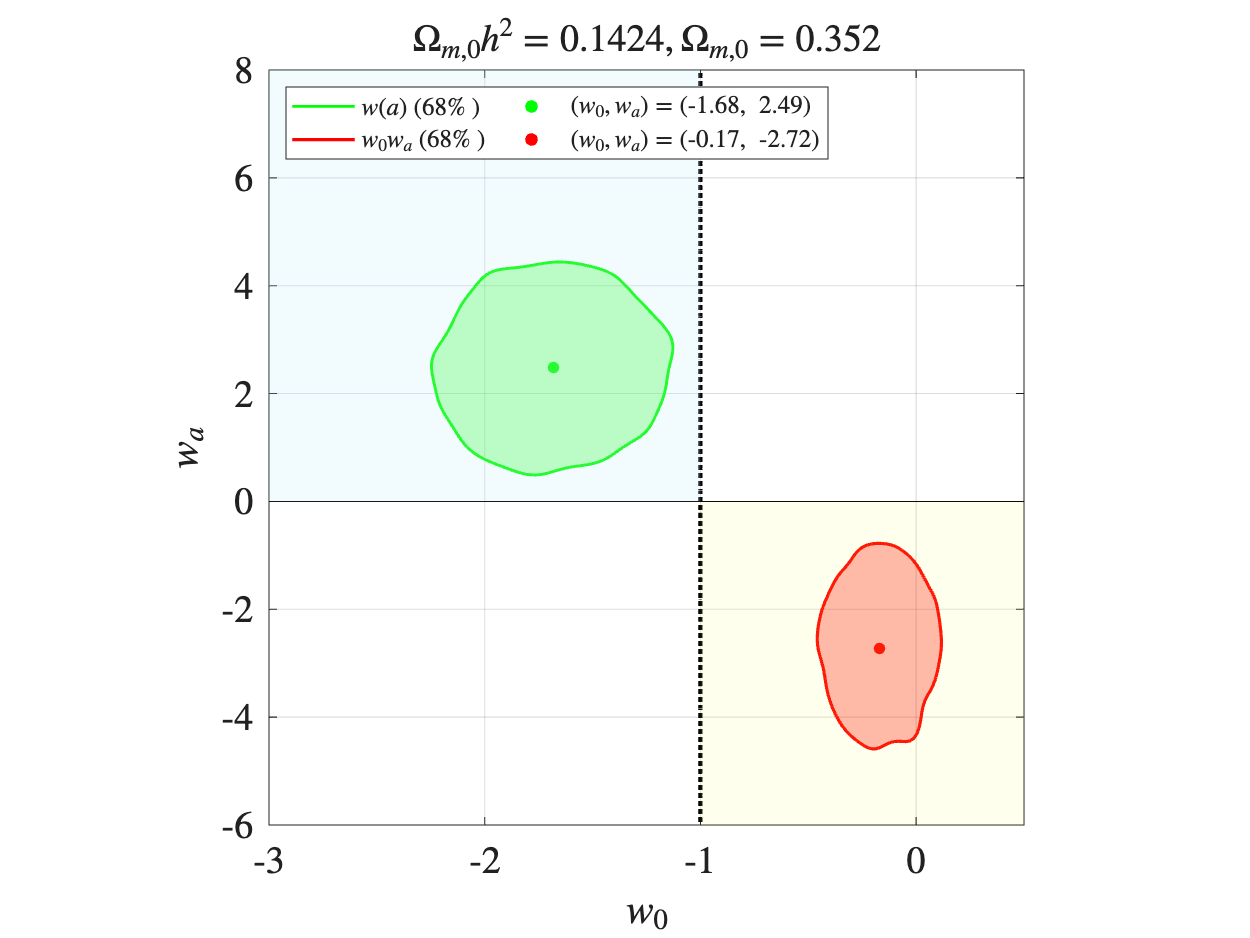}
    \includegraphics[width=0.48\linewidth]{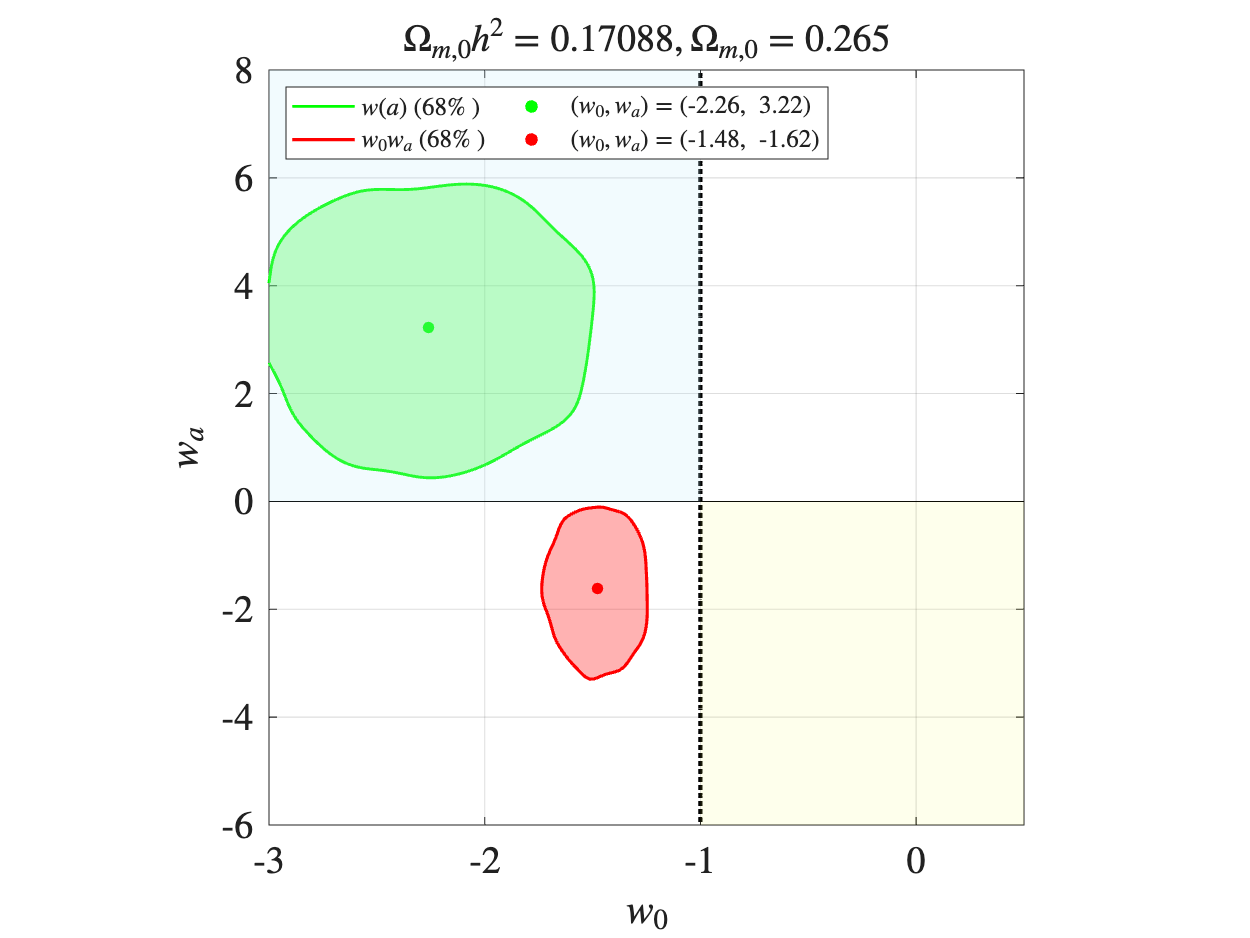}
    \includegraphics[width=0.48\linewidth]{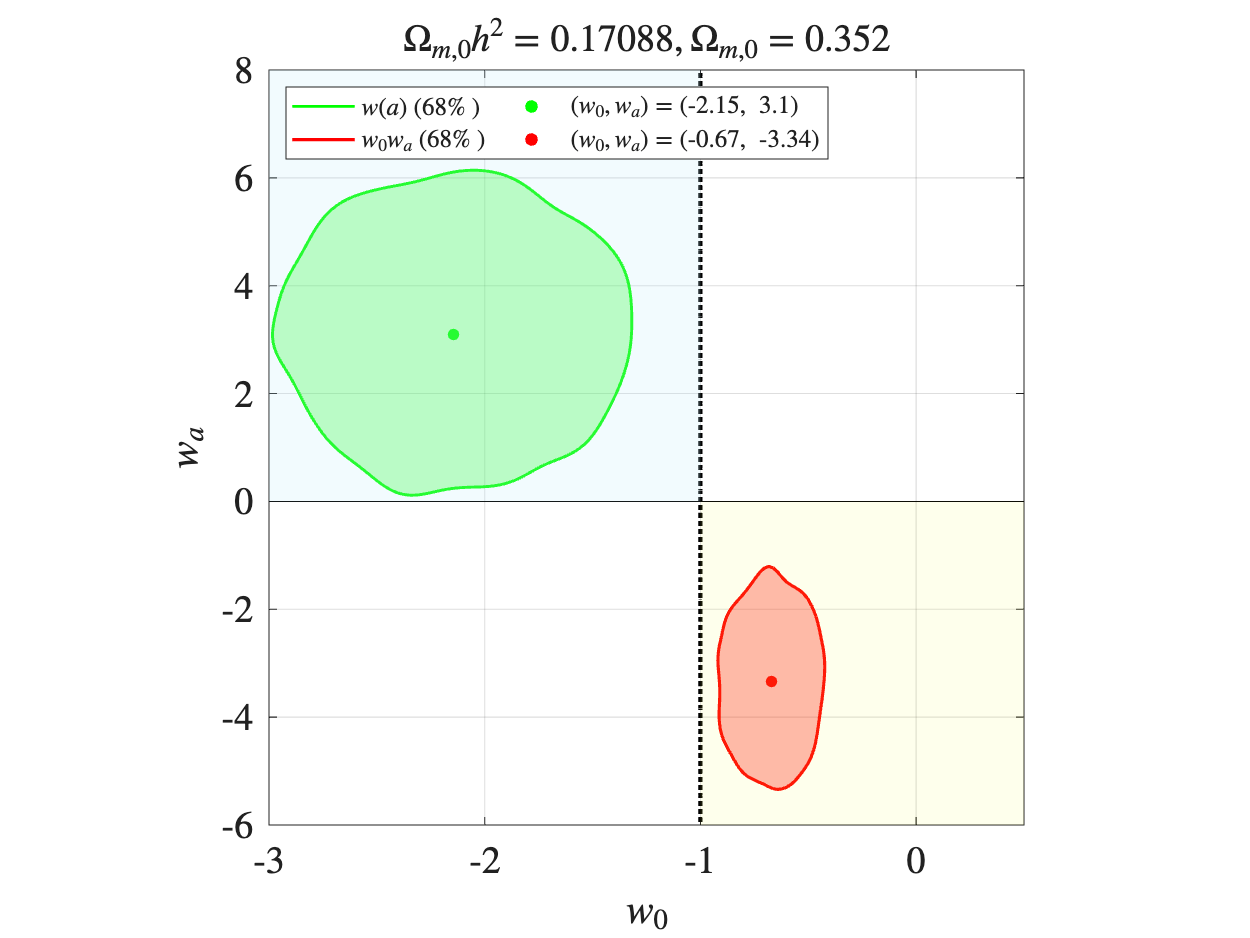}
    \caption{Sensitivity of $w_0w_a$ estimates in the $w_0w_a$ plane to the assumed values of $\Omega_{m,0}$ and $c_M$, where green and red refer to late and early linearization, respectively. Rows show sensitivity to constraint-preserving variations; columns show sensitivity to constraint-violating variations for three values of $c_M$. Increasing $c_M$ while keeping $\Omega_{m,0}$ constant moves both green and red to lower $w_0$ and higher (similar) $w_a$. In contrast, keeping $c_M$ constant while increasing $\Omega_{m,0}$ does not affect the output for green, but pushes the red island to the southeast with higher $w_0$ and lower $w_a$. These sensitivities are quantified in \ref{EQN_sensitivity} and \ref{EQN_sensitivity2}.}
    \label{fig:6panel}
\end{figure*}

Fig. \ref{fig:6panel} shows the sensitivity to changes in $c_M$ and $\Omega_{m,0}$. Notably, increasing $c_M$ shifts $w_0$ to lower values in early (red) and late (green) linearization. 
On the other hand, it increases $w_a$ of the green island, while it decreases the same of the red island. These sensitivities, though of opposite sign, 
are of the same order of magnitude.

Varying $\Omega_{m,0}$ while keeping $c_M$ constant has a different effect. 
As expected from \eqref{EQN_dw}, the green island remains fixed (rows in Fig. \ref{fig:6panel}) while
red noticeably drifts with $w_0$ increasing and $w_a$ decreasing.
We summarize Fig. \ref{fig:6panel} as follows:

\begin{equation}
\left(
\begin{matrix}
\frac{\partial w_0}{\partial \Omega_{m,0}} \\
\frac{\partial w_a}{\partial \Omega_{m,0}}
\end{matrix}\right)_g=
\begin{pmatrix}
-0.57\\
-0.69
\end{pmatrix},\hspace{0.2cm}
\left(\begin{matrix}
\frac{\partial w_0}{\partial \Omega_{m,0}} \\
\frac{\partial w_a}{\partial \Omega_{m,0}}
\end{matrix}\right)_r=
\begin{pmatrix}
7.93\\
-14.83
\end{pmatrix},
\label{EQN_sensitivity}
\end{equation}

\noindent where results for green are statistically consistent with zero. 
These results on $\Omega_{m,0}$-correlations are revealing: 
late linearization results remain stable, while early linearization appears dramatically unstable
quantitatively consistent with the correlations
(\ref{EQN_M0}-\ref{EQN_SM}) obtained from
mock data. The distinct behavior of the two
methods is expressed by the ratios:
\begin{equation}
    \left|\frac{(\partial w_0/\partial \Omega_{m,0})_r}{(\partial w_0/\partial \Omega_{m,0})_g}\right|=13.9   \hspace{0.2cm} \mbox{and} \hspace{0.2cm}\left|\frac{(\partial w_a/\partial \Omega_{m,0})_r}{(\partial w_a/\partial \Omega_{m,0})_g}\right|=21.5 .
    \label{EQN_sensitivity2}
\end{equation}

\noindent These ratios are anomalously large, 
indicating the strongly {non-commutative} behavior
of $w_0w_a$ estimation and linearization of
$w(a)$CDM.

\end{document}